\documentclass[fleqn]{scpma}
\usepackage[utf8]{inputenc}
\usepackage[T1]{fontenc}
\usepackage{ae,aecompl}
\usepackage{geometry}
\usepackage{pdflscape}

\usepackage{graphicx}
\usepackage{float}
\usepackage{amssymb}
\usepackage{amsmath}
\usepackage{relsize}
\usepackage{adjustbox}

\def\cvir{c_\mathrm{200m}}
\def\rvir{r_\mathrm{200m}}

\def\dispDM{\sigma_{\mathrm{DM}}}

\def\siglogM{\sigma_{\log M}}
\def\pimax{\Pi_{\mathrm{max}}}
\def\Mpch{\mathrm{Mpc}\,h^{-1}}
\def\Acon{\mathcal{A}_{\mathrm{con.}}}

\def\Qcen{\mathcal{Q}_{\mathrm{cen}}}
\def\Qsat{\mathcal{Q}_{\mathrm{sat}}}

\def\del8{\delta_8^g}
\def\Msun{M_\odot}

\def\Datvec1{\{w_p(r_p), n_g \}}
\def\Datvec2{\{w_p(r_p), f(\delta_8^g), n_g \}}
\def\Datvec3{\{ w_{p} (r_p | \tilde{\delta}_8^{g,i})\rvert_{i=1\mbox{-}5}, n_g \}}
\def\Datvec4{\{ w_{p} (r_p | \tilde{\delta}_8^{g,i})\rvert_{i=1\mbox{-}5}, f(\delta_8^g), n_g \}}

\begin{document}

\ensubject{subject}

\ArticleType{Article}
\SpecialTopic{SPECIAL TOPIC: }
\Year{2022}
\Month{January}
\Vol{60}
\No{1}
\DOI{10.1007/s11432-016-0037-0}
\ArtNo{000000}
\ReceiveDate{January 11, 2022}
\AcceptDate{April 6, 2022}

\title{Elucidating Galaxy Assembly Bias in SDSS}{Elucidating Galaxy Assembly Bias in SDSS}

\author[1]{Andr\'{e}s N. SALCEDO}{}
\author[2,3]{Ying ZU}{{yingzu@sjtu.edu.cn}}
\author[4]{Youcai ZHANG}{}
\author[5,6]{Huiyuan WANG}{}
\author[2,3,7]{\\Xiaohu YANG}{}
\author[4]{Yiheng WU}{}
\author[2,3,7]{Yipeng JING}{}
\author[8]{Houjun MO}{}
\author[1]{David H. WEINBERG}{}

\AuthorMark{Salcedo A. N.}

\AuthorCitation{Salcedo A. N., Zu Y., Zhang Y., et al.}

\address[1]{Department of Astronomy and Center for Cosmology and AstroParticle Physics, \\The Ohio State University, Columbus, OH 43210, USA}
\address[2]{Department of Astronomy, School of Physics and Astronomy, Shanghai Jiao Tong University, Shanghai 200240, China}
\address[3]{Shanghai Key Laboratory for Particle Physics and Cosmology, Shanghai Jiao Tong University, Shanghai 200240, China}
\address[4]{Key Laboratory for Research in Galaxies and Cosmology, Shanghai Astronomical Observatory, Shanghai 200030, China}
\address[5]{CAS Key Laboratory for Research in Galaxies and Cosmology, Department of Astronomy, \\University of Science and Technology of China, Hefei, Anhui 230026, China}
\address[6]{School of Astronomy and Space Science, University of Science and Technology of China, Hefei 230026, China}
\address[7]{Tsung-Dao Lee Institute, Shanghai Jiao Tong University, Shanghai 200240, China}
\address[8]{Department of Astronomy, University of Massachusetts Amherst, MA 01003, USA}

\abstract{We investigate the level of galaxy assembly bias in the Sloan
Digital Sky Survey (SDSS) main galaxy sample using ELUCID, a
state-of-the-art constrained simulation that accurately reconstructed the
initial density perturbations within the SDSS volume. On top of the ELUCID
haloes, we develop an extended HOD model that includes the assembly bias of
central and satellite galaxies, parameterized as $\Qcen$ and $\Qsat$,
respectively, to predict a suite of one- and two-point observables. In
particular, our fiducial constraint employs the probability distribution of
the galaxy number counts measured on $8\,\Mpch$ scales $N_8^g$ and the
projected cross-correlation functions of quintiles of galaxies selected by
$N_8^g$ with our entire galaxy sample. We perform extensive tests of the
efficacy of our method by fitting the same observables to mock data using
both constrained and non-constrained simulations. We discover that in many
cases the level of cosmic variance between the two simulations can produce
biased constraints that lead to an erroneous detection of galaxy assembly
bias if the non-constrained simulation is used. When applying our method to
the SDSS data, the ELUCID reconstruction effectively removes an otherwise
strong degeneracy between cosmic variance and galaxy assembly bias in SDSS,
enabling us to derive an accurate and stringent constraint on the latter.
Our fiducial ELUCID constraint, for galaxies above a stellar mass threshold
$M_*{=}10^{10.2}\,h^{-2}\,M_\odot$, is $\Qcen{=}{-}0.09\pm{0.05}$ and
$\Qsat{=}0.09\pm{0.10}$, indicating no evidence for a
significant~($>2\sigma$) galaxy
assembly bias in the local Universe probed by SDSS. Finally, our method
provides a promising path to the robust modelling of the galaxy-halo
connection within future surveys like DESI and PFS.}

\keywords{spatial distribution of galaxies, large scale structure of the Universe, Numerical simulation}

\PACS{98.62.Py, 98.65.Dx, 02.60.Cb }

\maketitle

\begin{multicols}{2}

\section{Introduction}

As one of the most successful models of the galaxy-halo connection
\cite{Wechsler_Tinker_rev_2018}, the halo occupation distribution~(HOD)
provides a convenient yet precise analytic framework for interpreting
modern galaxy surveys (e.g.~\cite{Jing_Mo_Borner_1998, Ma_Fry_2000,
Peacock_Smith_2000, Seljak_2000, Scoccimarro_et_al_2001, Berlind_2002,
Cooray_Sheth_2002, Yang_et_al_2003, vdBosch_et_al_2003b, Zheng_et_al_2005,
Cooray_2006, Mandelbaum_et_al_2006, Leauthaud_et_al_2012,
Zu_Mandelbaum_2015}).  In its simplest form, the standard HOD model assumes
that the galaxy occupation inside dark matter haloes depends solely on halo
mass. The physical basis for this assumption can be traced back to early
theories of structure formation (e.g.~\cite{Press_Schechter_1974,
Bardeen_et_al_1986, Kaiser_1986, Bond_et_al_1991, Bower_1991,
Mo_White_1996, Sheth_Mo_Tormen_2001}) in which the gravitational collapse
and spatial clustering of dark matter halos are determined by the peak
height in the initial density field.  Since then, various numerical studies
have robustly detected the so-called ``halo assembly bias'' effect using
cosmological N-body simulations --- halo clustering also depends on halo
properties~(e.g., concentration, formation time, spin, etc.) other than the
mass (e.g.~\cite{Sheth_Tormen_2004,Gao_2005,Harker_et_al_2006,
Wechsler_2006, Gao_White_2007, Jing2007, Wang_Mo_Jing_2007, Li_Mo_Gao_2008,
Faltenbacher_White_2010, Mao_Zentner_Wechsler_2018, Salcedo_2018,
Sato-Polito_et_al_2019, Xu_Zheng_2018, Johnson_et_al_2019}).  If galaxy
occupation inside haloes of the same mass varies with respect to a
secondary halo property that manifests strong halo assembly bias the
resulting ``galaxy assembly bias'' will make the clustering predictions of
a standard HOD model inaccurate; HOD prescriptions and other models of the
galaxy-halo connection would need to be modified accordingly. The potential
existence of galaxy assembly bias is thus an important source of systematic
uncertainty in cosmological interpretation of current and future galaxy
surveys~\cite{Croton_et_al_2007, Zu_et_al_2008, McCarthy_et_al_2019}.  In
this paper, we extend the standard HOD prescription to include both central
and satellite assembly biases, and we constrain the level of galaxy
assembly bias in the local Universe observed by the Sloan Digital Sky
Survey~(SDSS) using a novel combination of one and two-point statistics.

Observationally, many studies have attempted to detect galaxy assembly bias
directly from SDSS, but the results so far are inconclusive or even in
contradiction with one another. Using the SDSS Main Galaxy Sample, Zentner
et al.\ \cite{Zentner_et_al_2019} constrained galaxy assembly bias by
fitting the galaxy clustering of different volume-limited galaxy samples
with the galaxy assembly bias prescription of Hearin et al.\
\cite{Hearin_et_al_2016}. They found that the $M_r{<}-20$ sample favors
non-zero central assembly bias at roughly the $3\sigma$ level, while the
$M_r{<}-19$ sample favors non-zero satellite assembly bias at a weaker
level. Focusing on the central galaxies of SDSS
groups~\cite{Yang_et_al_2007}, Lin et al.\ \cite{Lin_et_al_2016} found
little difference between the large-scale clustering of early and
late-forming central galaxies that reside in haloes of the same weak
lensing mass (but see~\cite{Yang2006}).  However, applying a similar
detection philosophy to the more massive luminous red galaxy sample in the
SDSS-III Baryon Oscillation Spectroscopic Survey (BOSS;
\cite{Dawson_BOSS_et_al_2013}), Niemiec et al.\ \cite{Niemiec_et_al_2018}
claimed that the galaxies exhibit significant large-scale clustering
discrepancies when divided into two subsamples based on stellar
age~\cite{Montero-Dorta_et_al_2017}.

Aside from galaxy clustering, the observed large-scale correlation of
galaxy colours\footnote{For the small scale counterpart to this effect see
\cite{Weinmann_et_al_2006}.}(a.k.a ``galactic
conformity''~\cite{Kauffmann_et_al_2013}) had been regarded as potential
evidence of galaxy assembly bias \cite{Hearin_et_al_2015}. However, Zu and
Mandelbaum \cite{Zu_Mandelbaum_2018} found that the HOD framework of
\cite{Zu_Mandelbaum_2015} can successfully explain the large-scale galactic
conformity phenomenon, by combining the environmental dependence of the
halo mass function~\cite{Lemson1999, Faltenbacher2010, Zhang2014, Yang2017}
and the halo mass quenching model~\cite{Zu_Mandelbaum_2016}, without any
need for galaxy assembly bias. Other studies have reached similar
conclusions with different methods ~\cite{Sin_et_al_2017,
Tinker_et_al_2018, Calderon_et_al_2018}.  Furthermore, Alam et al.\
\cite{Alam_et_al_2019} studied the dependence of galaxy clustering on
density and geometric environment within SDSS. They found that the same
HOD+halo quenching model can also reproduce the colour-dependent clustering
of galaxies selected by large-scale overdensity as well as tidal
anisotropy, which N-body studies show to be one of the most important
drivers of halo assembly bias \cite{Paranjape_et_al_2018}.  Therefore, any
galaxy assembly bias effect appears to be weak  in comparison to the
environmental dependence of the halo mass function, which can be fully
accounted for in the standard HOD.

Since the SDSS Main Galaxy Sample probes a relatively small volume
(${\sim}(650\,h^{-1}\mathrm{Mpc})^3$;\cite{Zehavi_et_al_2011}) with a
significant presence of cosmic variance in the large-scale
structures~\cite{Chen_ELUCIDVI_et_al_2019}, at least some of the
inconsistencies among various detection results may arise from the fact
that the cosmic variance in the halo clustering could be misinterpreted as
evidence for galaxy assembly bias. In this paper, we mitigate such
confusion by utilizing ELUCID, a state-of-the-art constrained simulation
that accurately reconstructs the initial density perturbations of the SDSS
volume. ELUCID provides a high-fidelity input halo catalogue that
faithfully reproduces the large-scale environment of each {\it individual}
halo in SDSS, effectively removing the impact of cosmic variance on our
analysis.

Despite the lack of strong observational evidence of galaxy assembly bias,
a wide variety of models have been proposed to extend traditional methods
of galaxy-halo connection to allow different forms of its existence, from
the ``age-matching model'' that introduced maximum correlation between
galaxy colours and halo formation time~\cite{Hearin_and_Watson_2013}, to
the modifications to subhalo abundance matching proposed by Contreras et
al.\ \cite{Contreras_et_al_2021}, to various proposed modifications to the
HOD that introduce a secondary dependence of the occupation on a halo
internal property at fixed halo mass~\cite{Hearin_et_al_2016,
Yuan_et_al_2018}. Alternatively, many studies have looked for insights on
galaxy assembly bias from hydrodynamic simulations and semi-analytic
models, but its behavior differs among models depending on the exact
details of the galaxy formation recipe~\cite{Wang2013,
Chaves-Montero_et_al_2016, Artale_et_al_2018,Zehavi_et_al_2018,
Bose_et_al_2019,  Contreras_et_al_2019, Beltz-Mohrmann_et_al_2020,
Xu_Zheng_2020, Hadzhiyska_et_al_2021, Xu_et_al_2021}. Instead of choosing a
specific halo internal property as our proxy for galaxy assembly bias,
following Wibking et al.\ \cite{Wibking_et_al_2019} we adopt a more
general/agnostic approach by parameterising central and satellite assembly
biases as the variation of their halo occupation with the dark matter
overdensity defined within a radius of $8.0 \; h^{-1} \; \mathrm{Mpc}$ (see
also~\cite{McEwen_2018,Salcedo_et_al_2020,Xu_et_al_2021}).

To further enhance our constraining power, in addition to the commonly used
galaxy two-point correlation function $w_p(r_p)$, we also include the
probability density distribution of large-scale galaxy number counts
$p(N_8^g)$. Wang et al.\ \cite{Wang_et_al_2019} showed that the combination
of one and two-point galaxy statistics can greatly improve on constraints
from two-point statistics alone. Therefore, we anticipate that by combining
correlation functions with galaxy observables that incorporate the
one-point large-scale galaxy number count $N_8^g$ we can produce more
stringent constraints on galaxy assembly bias. Inspired by Abbas and Sheth
\cite{Abbas_Sheth_2007}, we also employ the galaxy number count dependent
correlation functions $w_p(r_p |N_8^g)$, in hopes of extracting
overdensity-dependent HODs in an analogous fashion to deriving global HOD
parameters from the overall galaxy clustering.

The paper is organised as follows. In the next section we describe the data
and simulations we use. In particular, section \ref{subsec:sdss} describes
the SDSS-DR7 stellar mass limited galaxy catalog, while section
\ref{subsec:simu} describes the Bolshoi and ELUCID simulations and the
methods of halo identification applied to both simulations to produce halo
catalogs. Section \ref{sec:mocks} provides a detailed description of our
HOD modelling methodology, including our extensions to the HOD to model
galaxy assembly bias. Section \ref{sec:obs} describes the galaxy clustering
statistics we utilize in our analysis.  Section \ref{sec:model} details the
methods of our Markov chain Monte Carlo (MCMC) analysis. We present the
results of fitting our data vectors to mock galaxy catalogs and SDSS data
in sections \ref{sec:mock} and \ref{sec:result} respectively. We summarize
our results in section \ref{sec:conc}.

\section{Data and Simulations}
\label{sec:data}

\subsection{SDSS Main Galaxy Sample}
\label{subsec:sdss}

We make use of the final data release of the SDSS I/II
\cite{Abazajian_SDSS_et_al_2009}, which contains the completed data set of
the SDSS-I and SDSS-II. In particular, we employ the Main Galaxy Sample
\cite{Strauss_SDSS_et_al_2002} data from the dr72 large-scale structure
sample bright0 of the `New York University Value Added Catalogue'
(NYU-VAGC), constructed as described in Blanton et al.\
\cite{Blanton_et_al_2005}. We apply the `nearest-neighbour' scheme to
correct for the 7\% of galaxies that are without redshifts due to fibre
collisions, and use data exclusively within the contiguous area in the
North Galactic Cap and regions with angular completeness greater than 0.8.

We employ the stellar mass estimates from the latest MPA/JHU value added
galaxy catalogue\footnote{https://home.strw.leidenuniv.nl/~jarle/SDSS/}.
The stellar masses were estimated based on fits to the SDSS photometry
following the methods of Kauffmann et al.\ \cite{Kauffmann_et_al_2003} and
Salim et al.\ \cite{Salim_et_al_2007}, and assuming the Chabrier
\cite{Chabrier_2003} Initial Mass Function (IMF) and the Bruzual and
Charlot \cite{Bruzual_Charlot_2003} Stellar Population Synthesis (SPS)
model.

\begin{figure*}[htb!]
\centering \includegraphics[width=1.0\textwidth]{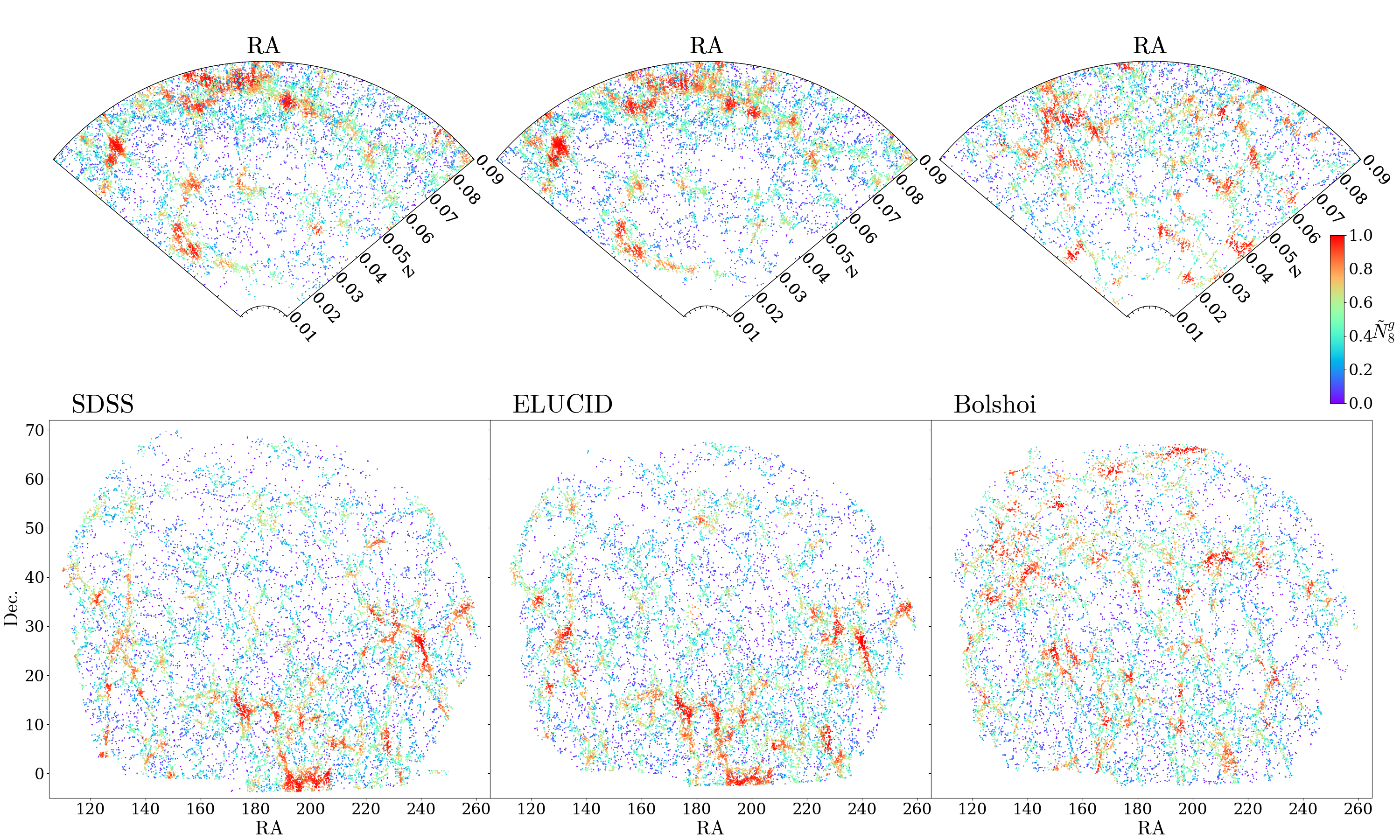}
    \caption{Redshift wedges (top, $\mathrm{Dec.} {<} 10.0^{\circ}$) and
    redshift
slices (bottom, $z = 0.08 - 0.09$) in SDSS (left), ELUCID (middle), and
Bolshoi (right). Points represent the positions of galaxies, colored by
their rank in the galaxy number count $\tilde{N}_{8}^{g}$ (described in
\ref{sec:galdens}). The striking similarity between SDSS and ELUCID
    demonstrates the efficacy of reconstruction in ELUCID, in sharp
    contrast
with Bolshoi, which is a random realization of $\Lambda\mathrm{CDM}$.}
\label{fig:footprint}
\end{figure*}

We select all the SDSS galaxies within the redshift range $z = [0.01,0.09]$
and with stellar mass above $M^\mathrm{min}_* = 10^{10.2} \; h^{-2} \;
\mathrm{M}_\odot$, yielding a sample of 82,824 galaxies in total. Based on
the stellar mass completeness limit estimated by Zu and Mandelbaum
\cite{Zu_Mandelbaum_2015}, we believe this cut in redshift produces a
galaxy sample that is roughly volume complete down to $M^\mathrm{min}_*$.
The volume of the resulting sample is ${\sim}(230 \, h^{-1} \,
\mathrm{Mpc})^3$. Note that by extending the maximum redshift of our sample
to $0.09$, we will include the ``Sloan Great
Wall''~\cite[SGW]{Gott_et_al_2005} at $z\sim 0.08$ in our analysis. As one
of the densest regions within the SDSS volume, galaxies in the SGW could
potentially reveal the strongest galaxy assembly bias signal within our
sample.

\subsection{Simulations: ELUCID vs. Bolshoi}
\label{subsec:simu}

One of the key ingredients of our fiducial model is the constrained
simulation ELUCID~\cite{Wang_ELUCIDIII_et_al_2016}, which is a $500^3 \;
h^{-3} \; \mathrm{Mpc}^3$ cubic box containing $3072^3$ particles with a
particle mass of $3.088 \times 10^8 \; h^{-1} \; \Msun$. The ELUCID
cosmology is based on the results of the WMAP5
\cite{Dunkley_WMAP5_et_al_2009, KomatsuWMAP_2011}: a flat Universe with
$\Omega_\kappa = 0$, $\Omega_{m,0} = 0.258$, $\Omega_{\Lambda,0} = 0.742$,
$h=0.72$, $n_s=0.96$, and $\sigma_8 = 0.80$. Below we will briefly describe
the ELUCID reconstruction, and refer to Wang et al.\
\cite{Wang_ELUCIDI_et_al_2014} for technical details of the method.

The ELUCID simulation reconstructs the density field of the nearby Universe
using the SDSS DR7 group catalog of Yang et al.\
\cite{Yang_et_al_2007}\footnote{For a detailed description of the group
finding algorithm see \cite{Yang_et_al_2005}.} and the Hamiltonian Markov
Chain~(HMC) Monte Carlo with Particle Mesh~(PM) dynamics algorithm of Wang
et al.\ \cite{Wang_ELUCIDI_et_al_2014}.  The reconstruction was performed
between $z{=}0.01$ and $0.12$, where the group catalog is roughly
volume-complete above $M_h{=}10^{12} \; h^{-1} \; \Msun$.  Following Wang
et al.\ \cite{Wang_et_al_2009b}, the redshift space distortions were
statistically removed to estimate the real space distances to the groups.
To reconstruct the present-day dark matter density field from the galaxy
groups, ELUCID employed a halo-domain reconstruction method which
partitions the local volume into a set of domains such that each domain
contains exactly one halo and every point within the domain is closer to
its halo than to any other using the distance metric $r / R_h$ where $R_h$
is the virial radius of the domain halo. Subsequently, pre-computed
simulated halo-matter profiles are used to produce a density profile within
each domain based on the mass of the domain halo.  Finally, the HMC+PM
method of Wang et al.\ \cite{Wang_ELUCIDI_et_al_2014} is applied to this
present-day density field to reconstruct the initial density field at
$z_\mathrm{ini} = 100$, which is then evolved to $z = 0$ using L-GADGET, a
memory optimized version of GADGET-2 \cite{Springel_GADGET2_2005}.

To compare to our fiducial results from ELUCID, we also use the publicly
available Bolshoi-Planck
simulation~\cite{Klypin_Bolshoi_et_al_2011},\footnote{https://www.cosmosim.org/cms/simulations/bolshoip/}
which is a $250^3 \; h^{-3} \; \mathrm{Mpc}^3$ cubic box with $2048^3$
particles, each $1.350 \times 10^8 \; h^{-1} \; \Msun$ in mass.  The
Bolshoi-Planck cosmology is based off of Planck 2016 \cite{Planck_2016},
with $\Omega_\kappa = 0$, $\Omega_{m,0} = 0.3071$, $\Omega_{\Lambda,0} =
0.6929$, $h=0.70$, $n_s = 0.96$, and $\sigma_8 = 0.82$. We will refer to
this simulation simply as ``Bolshoi'' for the rest of the paper.  For both
simulations, we use the spherical overdensity halo catalogues produced by
the standard Rockstar halo finder \cite{Behroozi_2013}. Our halo mass is
defined so that the average density enclosed within the halo radius is
$200$ times the mean background density of the Universe.

Despite the slight differences in cosmology and mass resolution, we
emphasize that the most important difference between these two simulations
is the density reconstruction --- ELUCID provides a dark matter density
field that faithfully reproduces the large-scale environment of each {\it
individual} group-sized halo observed in SDSS while Bolshoi does not.
Therefore, any difference in the constraints on the galaxy-halo connection
obtained using the two simulations should be driven primarily by cosmic
variance, i.e., by the difference between structure in the SDSS volume
modeled by ELUCID and the random realization of a periodic cube in Bolshoi.
By comparing results from Bolshoi and ELUCID we are able to determine the
extent to which cosmic variance can bias constraints and potentially
produce erroneous detections of galaxy assembly bias.

Figure \ref{fig:footprint} illustrates the effectiveness of the ELUCID
reconstruction. Each panel shows the distribution of galaxies color-coded
by a measurement of their large-scale galaxy number count $N_8^g$
(described in detail in \S~\ref{sec:galdens}), with red points existing in
the most dense regions and purple in the least dense regions. The left
panels show the distribution of the SDSS galaxies in our sample, while the
middle panels show the distribution of mock galaxies within the
SDSS-constrained region of ELUCID (note that this volume is not exactly the
same as in data), generated using the best-fitting parameters of our
fiducial constraint.  The right panels show the distribution of Bolshoi
mock galaxies populated using the same parameters as used for ELUCID. The
visual similarities between the SDSS and ELUCID panels are striking,
indicating that the ELUCID reconstruction, combined with our best-fitting
HOD parameters, provides a high-fidelity galaxy density field that closely
mimics the SDSS observation. While the ELUCID reconstruction is not perfect,
especially in the low density regions where few groups above $M_h{=}10^{12}
h^{-1} \mathrm{M_{\odot}}$ exist, the cosmic variance effect should be
greatly reduced in the intermediate and high-density regions where the
signal-to-noise of our measurements is the highest. In contrast, the
Bolshoi simulation exhibits a typical large-scale structure pattern formed
in $\Lambda\mathrm{CDM}$, but it clearly fails to capture the most
prominent structures in SDSS (e.g., the SGW). Naively, one might think that
the Bolshoi simulation is adequate for constraining galaxy assembly bias if
the galaxy sample is cut off at $z=0.075$ to avoid the SGW. However, the
strong cosmic variance effect cannot be mitigated by deliberately removing
the SGW from our analysis using Bolshoi, which could still produce a biased
constraint on galaxy assembly bias due to the lack of high-density
environments within the data.

\section{Populating Simulations with Galaxies}
\label{sec:mocks}

We populate simulated haloes with mock galaxies according to an extended
HOD framework. We will briefly describe the standard HOD component in
\S~\ref{sec:HOD} and introduce our extension for incorporating galaxy
assembly bias in \S~\ref{sec:assembly}.

\subsection{Halo Occupation Distribution}
\label{sec:HOD}

Following the standard practice (e.g. \cite{Kravtsov_et_al_2004,
Zheng_et_al_2005, Zehavi_et_al_2005, Zehavi_et_al_2011}), we separate
galaxies into centrals and satellites, so that for a stellar mass-threshold
sample haloes have an average probability $\langle
N_{\mathrm{cen}}(M_h)\rangle$ of hosting a central and the mean satellite
occupation $\langle N_{\mathrm{sat}}(M_h)\rangle$ is an increasing power
law in mass. We parametrize the two mean galaxy occupation numbers of
haloes as
\begin{align}
\left \langle N_{\mathrm{cen}}(M_h) \right \rangle &= \frac{1}{2} \left [ 1 + \mathrm{erf} \left ( \frac{\log M_h - \log M_{\mathrm{min}}}{\siglogM}\right ) \right ], \label{eq:cen_HOD} \\
\left \langle N_{\mathrm{sat}}(M_h) \right \rangle &= \left \langle N_{\mathrm{cen}} (M_h) \right \rangle \left ( \frac{M_h - M_0}{M_1} \right )^{\alpha}. \label{eq:sat_HOD}
\end{align}
where $M_\mathrm{min}$ refers to the characteristic mass to host a central (i.e. $\langle N_\mathrm{cen} (M_h) = 0.5 \rangle$),
$\siglogM$ is the width of the transition from $\langle N_\mathrm{cen} \rangle = 0.0$ to $\langle N_\mathrm{cen} \rangle = 1.0$,
$M_1$ is the normalization of the satellite-occupation power-law, $M_0$ truncates the satellite power-law, and $\alpha$ is the power-law index.

The number of centrals is sampled from a Bernoulli distribution. The number
of satellites placed into an individual halo is sampled from a Poisson
distribution centered on the mean satellite occupation. We place central
galaxies at the minimum of the potential of their host haloes. Satellites
are distributed according to a Navarro-Frenk-White (NFW; \cite{NFW_1997})
shape parametrized by halo concentration $\cvir = \rvir/r_s $ and
truncated at $\rvir$ the halo radius that encloses average density equal to
200 times the mean background density of the Universe. We measure halo
concentrations directly from the simulations and include an additional
parameter $\Acon$ to allow for the profiles of hosted satellite galaxies to
differ from the dark matter profiles of their host haloes
\begin{equation}
\cvir^\mathrm{gal} = \Acon \times \cvir^\mathrm{DM}.
\end{equation}

Finally, we include the effect of redshift space distortions in our mocks.
In both ELUCID and Bolshoi we measure the dark matter three dimensional
velocity dispersion of each of our halos $\dispDM$.  We assign central and
satellite velocities following the work of Guo et al.\
\cite{Guo_et_al_2015}. For each central galaxy, the velocity
$v_\mathrm{cen}$ relative to that of its host halo is drawn from an
exponential distribution,
\begin{equation}
p(v_{\mathrm{cen}} - v_h) = \frac{1}{\sqrt{2} \dispDM} \exp\left( - \frac{\sqrt{2} |v_{\mathrm{cen}} - v_h |}{\dispDM} \right),
\end{equation}
where $v_h$ is the relevant component of host halo velocity. For each satellite
galaxy we draw the velocity components relative to that of the host from a
Gaussian distribution with dispersion $\dispDM$. Guo et al.\ \cite{Guo_et_al_2015} include parameters to model central and satellite velocity bias, but since we have found these parameters to have very little effect on our observables we set them to
zero and unity respectively. We displace the positions of our mock galaxies using the line-of-sight components of the assigned velocities. In ELUCID we assume the position of the observer is at Earth, and in Bolshoi we adopt the $z$-axis as the line-of-sight
direction.

\subsection{Modeling Galaxy Assembly Bias}
\label{sec:assembly}

\begin{figure*}[htb!]
\centering \includegraphics[width=1.0\textwidth]{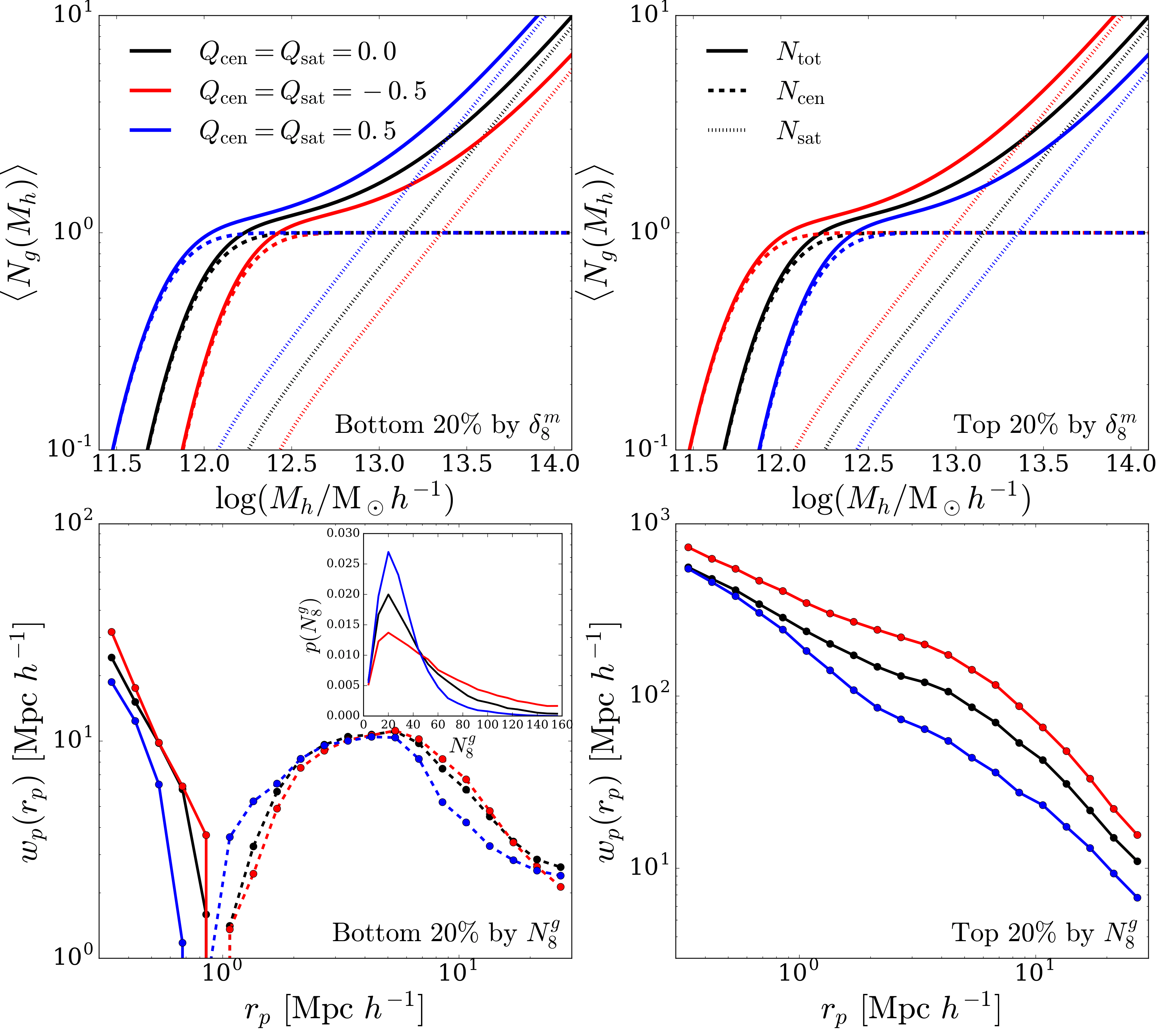}
    \caption{{\it Top panels}: HODs of galaxies in the bottom
     ($\tilde{\delta}^{m}_8{<}0.20$, left) and top
     ($\tilde{\delta}^{m}_8{>}0.80$, right)
      quintiles of matter overdensity rank
    $\tilde{\delta}^m_8$, for three galaxy assembly bias models with
    $\Qcen{=}\Qsat{=}0.0$~(black), $\Qcen{=}\Qsat{=}-0.5$~(red), and
    $\Qcen{=}\Qsat{=}0.5$~(blue), respectively. In both panels we show the
    total HOD (solid) along with central (dashed) and satellite (dotted)
    contributions. {\it Bottom panels}: Projected cross-correlation
    functions of the bottom~(left) and top~(right) galaxy quintiles of the
    {\emph{galaxy}} number count rank $\tilde{N}^g_8$ (described in
    \S~\ref{sec:galdens}) with the global galaxy sample for the three
    models of galaxy assembly bias.  In the bottom left panel the dashed
    lines correspond to negative values of the correlation function.
    Additionally we show the PDF of $N_8^g$ in the inset of the bottom left
    panel for the three models.}
\label{fig:hod}
\end{figure*}

For the purpose of our analysis, galaxy assembly bias refers to the
potential dependence of the galaxy occupation on properties other than halo
mass. Properties that are often considered in the literature include
internal properties like halo formation time and concentration, but this
definition does not exclude environmental properties such as the large
scale overdensity or tidal anisotropy. When such a secondary property
exhibits a halo assembly bias signal at fixed mass, the clustering of
galaxies will be modified from the standard HOD prediction on both small
and large scales.  In particular, in the absence of galaxy assembly bias,
the linear galaxy bias $b_g(M_h)$ at fixed halo mass $M_h$ is equal to the
linear halo bias $b_h(M_h)$. However, a galaxy assembly bias with respect
to halo property $x$ will introduce some covariance between
$N_\mathrm{gal}(x|M_h)$ and $b(x| M_{h})$, producing a linear galaxy bias
$b_g^{x}(M_h)$ that is modified from $b_g(M_h)$
\begin{align}
    b_g^{x}(M_h) =& \sum_{i=0}^{N_h} N_{\mathrm{gal}} (x_i \mid M_{h})  b(x_i \mid M_{h}) \,
    /\, \sum_{i=0}^{N_h} N_{\mathrm{gal}}( x_i \mid M_{h})\nonumber \\
    = & \langle N_{\mathrm{gal}}(x \mid M_{h})  b(x \mid M_{h})\rangle \, / \,
    \langle N_{\mathrm{gal}}(M_h)\rangle \nonumber \\
    = & b_g(M_h) + \mathrm{cov}(N_{\mathrm{gal}}, b \mid M_h)\,/\, \langle
    N_{\mathrm{gal}}(M_h)\rangle,
\end{align}
where $\mathrm{cov}(N_{\mathrm{gal}}, b \mid M_h)$ is the covariance
between galaxy occupation $N_{\mathrm{gal}}(x, M_h)$ and $b(x, M_h)$ at
$M_h$, and $\langle N_{\mathrm{gal}}(M_h)\rangle$ is the average galaxy
occupation at $M_h$. If we further assume the galaxy occupation follows the
Poisson distribution, which is an accurate approximation for high-occupancy
haloes, the extra term on the right can be expressed by
\begin{align}
    \Delta b_g =& b_g^{x}(M_h) - b_g(M_h) \nonumber \\
    =& \rho_{cc}\,\sigma(b \mid M_h)\,/\, \sqrt{\langle
    N_{\mathrm{gal}}(M_h)\rangle},
    \label{eqn:bsca}
\end{align}
where $\rho_{cc}$ and $\sigma(b \mid M_h)$ are the cross-correlation
coefficient and the scatter in halo bias at $M_h$, respectively. On small
scales, galaxy assembly bias will increase the second moment of the galaxy
occupation $\langle N_\mathrm{gal}^2 \mid M_h \rangle$ while minimally affecting
the mean occupation, thereby boosting galaxy clustering in the 1-halo regime
by adding more satellite-satellite pairs \cite{Berlind_2002}.

It is unclear which halo internal property is most responsible for galaxy
assembly bias if it exists. A variety of studies have examined galaxy
assembly bias within hydrodynamic simulations and semi-analytic models and
produced valuable insights (e.g. \cite{Wang2013, Chaves-Montero_et_al_2016,
Artale_et_al_2018, Zehavi_et_al_2018, Bose_et_al_2019,
Contreras_et_al_2019, Beltz-Mohrmann_et_al_2020, Xu_Zheng_2020,
Hadzhiyska_et_al_2021, Xu_et_al_2021}), but as of yet we lack a physical
and observational understanding of galaxy assembly bias. Although halo
formation time and concentration are some of the most commonly-used
secondary variables for modeling galaxy assembly bias (e.g.
\cite{Hearin_et_al_2016, Zentner_et_al_2019, Wang_et_al_2019}), there is no
observational evidence suggesting that one variable is more
physically-motivated than another. Therefore, we will take a more agnostic
approach and look for a generic variable that is mathematically more
convenient. According to Equation~\ref{eqn:bsca}, we would want a variable
that exhibits a very strong (anti-)correlation with halo bias, i.e.,
$|\rho_{cc}|\rightarrow 1$.  The obvious candidate variable is the halo
bias itself defined on the basis of individual haloes, i.e., the
large-scale dark matter overdensity. In particular, we implement a modified
version of the parameterization of Wibking et al. \cite{Wibking_et_al_2019}
(see also \cite{McEwen_2018, Salcedo_et_al_2020, Xu_et_al_2021}).  Wibking
et al.\ \cite{Wibking_et_al_2019} allow $M_\mathrm{min}$ to vary on a
halo-by-halo basis according to the large scale matter-overdensity measured
on $8\,\Mpch$ scales $\delta_8^m$. This environmental dependence is
written as
\begin{equation}
\log M_{\mathrm{min}} = \log M_{\mathrm{min},0} +  \Qcen \left ( \tilde{\delta}_8^m - 0.5 \right ), \label{eq:Qcen}
\end{equation}
where $\Qcen$ expresses the strength of the dependence of $M_\mathrm{min}$
on environment and $\tilde{\delta}_8^m \in [0,1]$ is the normalized rank of
$\delta_8^m$ within a narrow mass bin. Using the {\emph{rank}} of
$\delta_8^m$ has the advantage of being less sensitive to the particular
choice of $8\,\Mpch$.  When $\Qcen$ is positive haloes in dense
environments are less likely to host centrals, and vice versa when it is
negative. The case of $\Qcen = 0.0$ corresponds to having no assembly bias.
Similar to Xu et al.\ \cite{Xu_et_al_2021} we extend this parameterization
by allowing $M_1$ to also vary according to environment,
\begin{equation}
\log M_{1} = \log M_{1,0} + \Qsat \left ( \tilde{\delta}_8^m - 0.5 \right ). \label{eq:Q1}
\end{equation}

In the top two panels of Figure~\ref{fig:hod}, we illustrate the effect of
the two galaxy assembly bias parameters on the total (solid lines), central
(dashed) and satellite (dotted) halo occupations for galaxies selected by
their large scale matter overdensity $\delta_8^m$.  We plot the predicted
occupation for $\Qcen{=}\Qsat{=}0.0$ (black lines), $\Qcen{=}\Qsat{=}-0.5$
(red) and $\Qcen{=}\Qsat{=}0.5$ (blue), respectively, while keeping all the
other HOD parameters fixed. The top left panel shows variations of the HOD
for galaxies in underdense environments while the top right panel shows
variations for those in overdense environments. By design, when
$\Qcen{=}\Qsat{=}0$ there is no variation in HOD for different
environments.  However, when $\Qcen{=}\Qsat{=}-0.5$ both $M_\mathrm{min}$
and $M_1$ are increased (decreased) in underdense (overdense) environments.
This has the effect of boosting the galaxy occupation of haloes of a given
mass that are in overdense environments. When $\Qcen{=}\Qsat{=}0.5$ we
observe the opposite behavior, with both $M_\mathrm{min}$ and $M_1$
decreased (increased) in underdense (overdense environments).

\section{Observables}
\label{sec:obs}

\subsection{Galaxy Clustering}

We compute the projected galaxy correlation functions $w_p$ by integrating
the real-space correlation function along the line of sight
\begin{equation}
w_{p}(r_p) = 2 \int_{0}^{\pimax} \xi(r_p,\pi) \; d\pi,
\end{equation}
where $r_p$ and $\pi$ are the projected pair separation and the line-of-sight distance, respectively, and $\xi$ is the real-space isotropic correlation function.  We compute $\xi$ with the Natural estimator
\begin{equation}
\xi (r) = \frac{DD(r)}{RR(r)} - 1,
\end{equation}
where $DD(r)$ and $RR(r)$are the number of galaxy-galaxy and random-random
pairs with pair separation $r$, respectively. When computing $w_{p}$ in the
ELUCID and SDSS volumes we compute the random-random term using random
catalogs constructed with the same geometry as ELUCID and SDSS
respectively. When computing $w_{p}$ in Bolshoi we analytically compute the
random-random term as $RR{=}2 \pi \pimax (r_{p,\mathrm{max}}^2{-}r_{p,
\mathrm{min}}^2) ( N_g^2 / L^3_\mathrm{box})$.

We choose to set the integration limit $\pimax = 30\,h^{-1}\,\mathrm{Mpc}$.
This value is chosen to balance the need to have a large enough $\pimax$ to
mitigate the uncertainties in our modelling of the redshift-space
distortions and to suppress noise arising from uncorrelated structure along
the line of sight. We calculate $\xi(r_p, \pi)$ using {\sc{corrfunc}}
\cite{Sinha_2017} in 20 logarithmically-spaced bins from
$r_{p,\mathrm{min}}{=}0.3 \, h^{-1} \, \mathrm{Mpc}$ to
$r_{p,\mathrm{max}}{=}30.0 \, h^{-1} \, \mathrm{Mpc}$ and in
$1.0\,h^{-1}\,\mathrm{Mpc}$-wide bins in $\pi$, respectively.

\subsection{Galaxy Number Counts}
\label{sec:galdens}

To enhance the constraining power of our HOD model with ELUCID, we
supplement the two-point galaxy statistic $w_{p}$ with a one-point
statistic --- the probability density distribution~(PDF) of galaxy number
counts measured on $8\,\Mpch$ scales around each galaxy $N_8^g$ . When
computing $N_8^g$ for galaxies in the SDSS-constrained volume of ELUCID we
include neighbour galaxies outside of the SDSS-constrained volume. Although
this technically means our measurement of $N_8^g$ in ELUCID includes
contributions from non-constrained regions, the number of galaxies affected
is small.

Motivated by the results of Abbas and Sheth \cite{Abbas_Sheth_2007} and
Alam et al.\ \cite{Alam_et_al_2019}, we use the quantity $N_{8}^g$ to
define count-selected subsamples of galaxies and compute their clustering
statistics. Specifically, we divide our galaxies into quintiles of
$N_{8}^g$ and compute their projected cross-correlation functions with the
global sample. We denote these correlation functions as $w_{p}(r_p |
\tilde{N}_{8}^{g,i})$, where $\tilde{N}_{8}^{g,i}$ refers to the i-th
quintile selected by $N_8^g$.  In addition to the global clustering $w_{p}$
and the probability distribution of galaxy overdensity $p (N_{8}^{g})$ we
use these count-selected quintile cross-correlation functions to constrain
assembly bias. We anticipate that the clustering of the most extreme
over/underdense regions should be particularly sensitive to galaxy assembly
bias because it is in those regions that the HOD is most significantly
modified from the standard form if galaxy assembly bias exists (see
equations \ref{eq:Qcen} and \ref{eq:Q1}).

\subsection{Impact of Galaxy Assembly Bias on Observables}

The bottom two panels of Figure~\ref{fig:hod} illustrate the effects of the
galaxy assembly bias parameters on our observables. The bottom left panel
shows the variation among the cross-correlation functions between the
global sample and the bottom $20\%$ of galaxies selected by $N_8^g$ as
predicted by the three galaxy assembly bias models shown in the top left
panel.  Similarly, the bottom right panel shows the cross-correlation
variation for the top  $20\%$ of galaxies selected by $N_8^g$, with their
HODs shown in the top right panel. In both sets of panels, black lines show
results for $\Qcen{=}\Qsat{=}0.0$~(i.e., zero galaxy assembly bias), red
lines show results for $\Qcen{=}\Qsat{=}-0.5$, and blue lines for
$\Qcen{=}\Qsat{=}0.5$. In each case we keep all other HOD parameters fixed.
Note that these variations in $\Qcen$ and $\Qsat$ are exaggerated compared
to the values allowed by the data for illustrative purposes. We also show
the effects of assembly bias on the PDF of galaxy number count $p(N_8^g)$
in the inset of the bottom left panel.

For the galaxies in the bottom quintile of $N_8^g$, the correlation
functions are negative at large scales and we plot the amplitude of these
negative values with dashed lines.  On small scales, the cross-correlation
effectively measures the 1-halo term of the typical host halo. For the
$\Qcen{=}\Qsat{=}0.5$ case, the galaxies on average live in haloes of much
lower mass than those in the $\Qcen{=}\Qsat{=}-0.5$ case, producing a
significantly lower clustering on scales below $3\,h^{-1}\,\mathrm{Mpc}$.
On large scales, the $\Qcen{=}\Qsat{=}0.5$ case exhibits a stronger
cross-correlation (i.e., less negative). This is consistent with the inset
panel where the average $N_8^g$ of the bottom quintile in the
$\Qcen{=}\Qsat{=}0.5$ case is higher than the other two, because the model
places more galaxies in low-density regions.

The effects of $\Qcen$ and $\Qsat$ on the clustering of the highest galaxy
quintile selected by $N_8^g$ are shown in the bottom right panel of Figure
\ref{fig:hod}. In the case of $\Qcen{=}\Qsat{=}-0.5$, the clustering of
galaxies in the top $N_8^g$-quintile is boosted at all scales. At large
scales this is because the occupation of haloes in overdense environments
has been boosted.  On small scales, since haloes that host galaxies in
high-$N_8^g$ environments are generally more massive, and the number of
satellite pairs at fixed halo mass is higher compared to the
zero-assembly-bias mock, the galaxy clustering is also strongly boosted.
For the $\Qcen{=}\Qsat{=}0.5$ galaxies, the clustering signal decreases at
all scales relative to $\Qcen{=}\Qsat{=}0.0$. However, the difference is
the least significant at the smallest scales of $r_p$. This is because at
these small scales the effect of positive $\Qsat$ on $\langle
N_\mathrm{gal}^2 | M \rangle$ also acts to boost the number of satellite
pairs, which compensates for the impact from having a lower average halo
occupation in dense regions. At large scales the clustering is influenced
only by the latter effect, and therefore the effect of setting
$\Qsat{=}\Qcen{=}0.5$ is opposite in sign to that of setting
$\Qsat{=}\Qcen{=}-0.5$, albeit with similar amplitude.

Galaxy assembly bias also modifies the relative fraction of the low vs.
high overdensity environments, by shifting the galaxy population into halos
in low or high density regions. The inset of the bottom left panel of
Figure~\ref{fig:hod} shows such effects on the PDF of $N_8^g$ from setting
non-zero values of $\Qcen$ and $\Qsat$.  When $\Qcen{=}\Qsat{=}-0.5$ the
PDF is broadened with the density increasing at large values of $N_8^g$
relative to the PDF for $\Qcen{=}\Qsat{=}0.0$. This increase is driven by
the fact that negative values of $\Qcen$ and $\Qsat$ cause highly clustered
haloes to have higher occupations which increases the overall values of
$N_8^g$. This effect is mitigated however by the fact that the PDF has an
explicit integral constraint. The effect of setting $\Qcen{=}\Qsat{=}0.5$
on the halo occupation is the opposite, causing less clustered haloes to
have higher occupations for their mass. This causes the PDF to narrow and
decreases the number of galaxies with high $N_8^g$.

To summarize, Figure~\ref{fig:hod} demonstrates that, in the absence of
other systematic uncertainties, the combination of the PDF of $N_8^g$ and
the cross-correlation functions of galaxies selected by $N_8^g$ is a very
sensitive probe of galaxy assembly bias. Any deviation of $\Qcen$ and
$\Qsat$ from zero will generate coherent variation in the HOD between high
and low density regions of the Universe. This galaxy assembly bias effect
in the HOD, coupled with the environmental modulation of the halo mass
function, produces a unique set of phenomena in the overdensity
distribution and density-dependent cross-correlation functions that can
potentially be detected within the SDSS data.

\subsection{Distinguishing Cosmic Variance and Galaxy Assembly Bias}

\begin{figure*}[htb!]
\centering \includegraphics[width=0.80\textwidth]{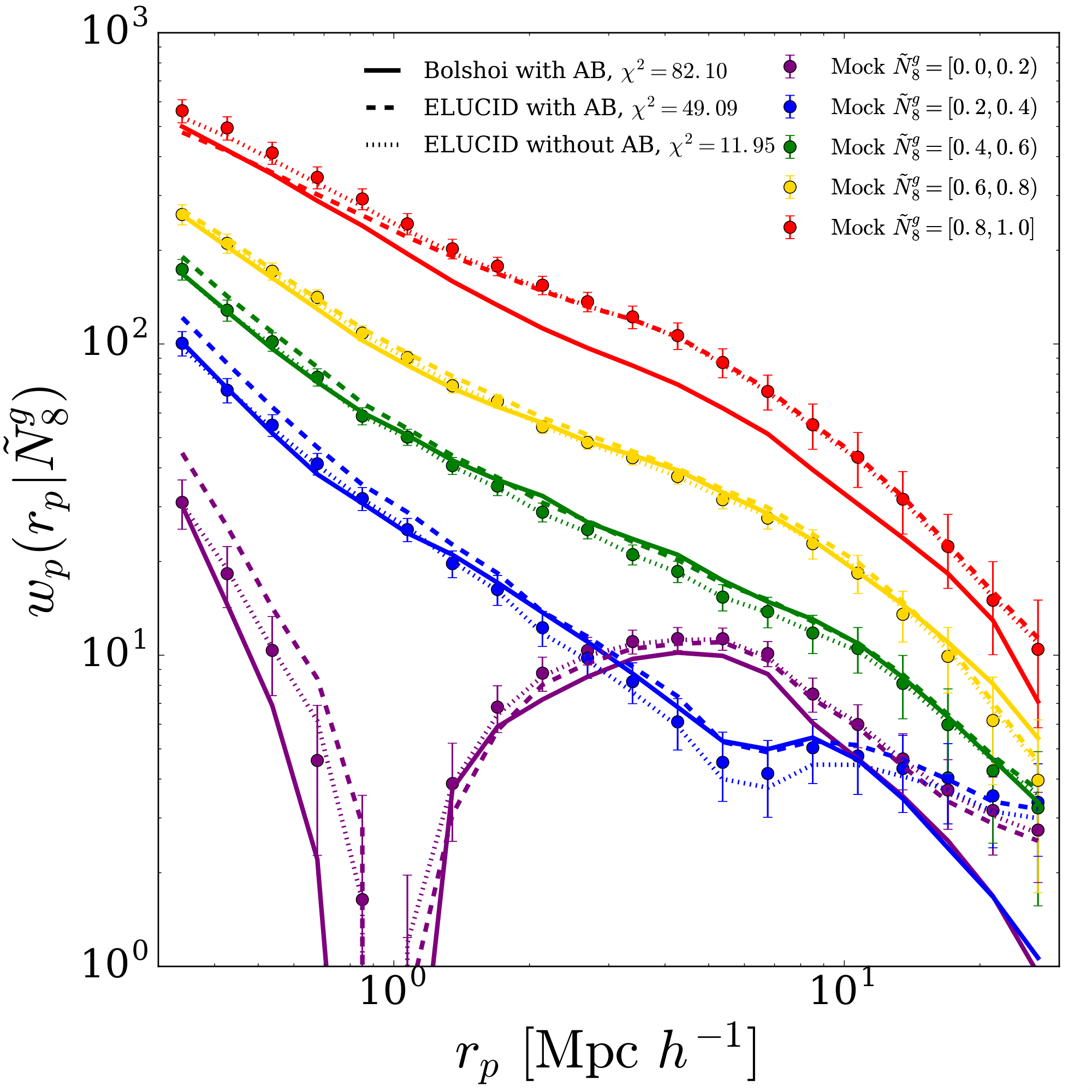}
\caption{
    Impact of cosmic variance on the projected cross-correlation functions
    between mock galaxies in number count $N^g_8$ quintiles and the global
    galaxy sample. Circles with errorbars show the measurements from an
    ELUCID galaxy mock built without galaxy assembly bias, while dotted
    lines show the results of a consistency test of fitting to this ELUCID
    galaxy mock using ELUCID halos~($\chi^2{=}11.95$).  Solid lines show
    the predictions from the best--fitting HOD model inferred from fitting
    to the mock data~(circles) using Bolshoi halos, which results in a
    false detection of galaxy assembly bias with $\Qcen{=}-0.15$ and
    $\Qsat{=}0.18$. Dashed lines show the results of directly applying the
    same HOD with non-zero values of $\Qcen$ and $\Qsat$ to ELUCID halos
    without any fitting. The false detection, along with the significant
    improvement in the goodness of fit from the solid~($\chi^2{=}82.10$) to
    the dashed lines~($\chi^2{=}49.09$), illustrates the vital importance
    of mitigating cosmic variance when constraining galaxy assembly bias.}
    \label{fig:cosmvar}
\end{figure*}

Cosmic variance can also significantly affect the measurements of one and
two-point statistics in the SDSS volume, including the stellar mass
function \cite{Chen_ELUCIDVI_et_al_2019} and the two-point correlation
function of the faint red galaxies \cite{Xu_et_al_2016}. Therefore, a
slight tendency of SDSS galaxies to reside in high~(low) density regions
due to cosmic variance could be misinterpreted as evidence for
negative~(positive) values of $\Qcen$ and $\Qsat$.

To mitigate the confusion due to cosmic variance, we utilize the ELUCID
constrained simulation as an input to our fiducial HOD analysis. As shown
in Figure \ref{fig:footprint}, ELUCID reproduces most of the specific
structure observed in the local SDSS volume, in sharp contrast with the
Bolshoi simulation that does not. Because Bolshoi and ELUCID have
reasonably similar cosmologies this difference is primarily driven by
cosmic variance. It is possible for this cosmic variance to bias
constraints on assembly bias (or other HOD parameters) obtained from
fitting our galaxy observables. This is particularly the case for the
$N_8^g$-selected correlation functions, because they are very sensitive to
the extremely low or high-density structures. For example, Zehavi et al.\
\cite{Zehavi_et_al_2011} identified that the Sloan Great Wall is the most
significant cosmic variance effect on their measurements of the global
$w_p$, and we thus expect a much stronger impact on the $w_p$ of galaxies
in the top quintile of $N_8^g$.

Figure~\ref{fig:cosmvar} shows a more quantitative comparison of the cosmic
variance and galaxy assembly bias effects on the five $N_8^g$-selected
correlation functions. Circles with errorbars are the measurements from a
mock galaxy sample constructed using ELUCID without galaxy assembly bias.
Using this measurement as our mock observables, we derive constraints
$\Qcen{=} -0.15^{+0.06}_{-0.06}$ and $\Qsat{=}0.18^{+0.17}_{-0.15}$ with
our extended HOD model using the Bolshoi simulation. The solid lines are
the predictions from the best-fitting model with $\Qcen{=}-0.15$ and
$\Qsat{=}0.18$.  As a consistency check, dotted lines indicate the
prediction from the best-fitting HOD model that uses ELUCID haloes as
input~(a fit that favors no assembly bias by recovering the true values of
$\Qcen{=}\Qsat{=}0.0$).  Clearly, the Bolshoi best-fit does a very poor job
describing the $w_p$ of galaxies in the top~(red) and bottom~(purple)
quintiles, largely due to cosmic variance between Bolshoi and ELUCID.
However, once we apply the Bolshoi best-fit as is directly to the ELUCID
haloes, the predictions, indicated by the dashed lines, provide a
reasonably good description of the data points without any fitting.  The
significant reduction of $\chi^2$ from $82.10$ to $49.09$ suggests that the
cosmic variance effect is the main hurdle to obtaining a satisfactory
goodness of fit for the HOD model, especially for galaxies in the highest
density regions. Because of cosmic variance, the Bolshoi fit spuriously
favors a non-zero value of $\Qcen$ at $2.5\sigma$ significance, and even
with the freedom allowed by assembly bias it cannot achieve a good match to
the data. The ELUCID mock with incorrect parameters reproduces the data
better but is nonetheless distinguishable at high significance from the
correct model.

One can argue that instead of describing the Bolshoi fit results as biased,
we should simply say that our observational errors are under-estimated.  To
overcome such shortcoming, we can either compute the covariance matrix from
a large number of simulations with different initial conditions, or base
our analysis on a constrained simulation like ELUCID that faithfully
reproduces the large-scale structure in the SDSS volume.  However, the
former approach is computationally prohibitive, while the latter approach,
which we adopt in this paper, has the added benefit that we can estimate
the covariance matrix via jackknife re-sampling of the data~(see
\S\ref{app:cov}), as is commonly done by similar analyses in the
literature, whether it be HOD modelling or detecting galaxy assembly bias.

\section{Gaussian Likelihood Model}
\label{sec:model}

In both our mock tests and our data constraints we consider four different
combinatorial data vectors,
\begin{enumerate}
\item The combination of projected galaxy two-point function and galaxy number
density $\{w_{p}(r_p), n_g\}$.
\item The combination of projected galaxy two-point function, PDF of
galaxy overdensity, and galaxy number density $\{w_{p}(r_p), p(N_8^g),
n_g\}$.
\item The cross-correlation functions between galaxies in quintiles of $N_8^g$
and the global sample, and the galaxy number density $\{ w_{p} (r_p |
        \tilde{N}_8^{g,i})\rvert_{i{=}1\mbox{-}5}, n_g \}$.
\item The cross-correlation functions between galaxies in quintiles of
$N_8^g$ and the global sample, the PDF of galaxy overdensity, and the
galaxy number
	density $\{ w_{p} (r_p | \tilde{N}_8^{g,i})\rvert_{i{=}1\mbox{-}5},
p(N_8^g), n_g \}$. This is our fiducial data vector.
\end{enumerate}

To constrain HOD and galaxy assembly bias parameters we fit to these four
data vectors measured from either SDSS galaxies or mock galaxies
constructed using ELUCID haloes. In each case we use the appropriate
covariance matrix described in \ref{app:cov} and assume $5\%$ Gaussian errors on $n_g$, uncorrelated with errors in other observables. We assume a Gaussian likelihood
model, $\mathcal{L} \propto e^{-\chi^2/2}$, where
\begin{equation}
\chi^2 =  \frac{ (n_g^\mathrm{mock} - n_g)^2}{ (0.05 \times n_g)^2 } +  \sum_{i,j} \Delta \mathcal{D}(x_i) \left[ C^{-1} \right]_{ij} \Delta \mathcal{D}(x_j),
\end{equation}
where $C^{-1}$ is the inverse covariance matrix and $\Delta
\mathcal{D}(x_i){\equiv}\mathcal{D}^\mathrm{mock}(x_i)-\mathcal{D}(x_i)$ is
the difference between mock and measured data vector $\mathcal{D}$. We
measure $w_{p}$, and $w_{p} (r_p |
\tilde{N}_8^{g,i})\rvert_{i{=}1\mbox{-}5}$ in 20 logarithmically spaced
bins from $0.3$ to $30.0 \, h^{-1} \, \mathrm{Mpc}$ with
$\pimax{=}30.0\,h^{-1}\,\mathrm{Mpc}$. We measure $p(N_8^g)$ in 20 linearly
spaced bins from $N_8^g{=}0.0$ to $N_8^g{=}160.0$. When measuring any of
our data vectors in ELUCID we only use galaxies within the SDSS constrained
volume and within redshift range $0.01{\leq}z{\leq}0.09$. This choice is
made to match to the volume of our SDSS data sample. When using Bolshoi we
use the entire volume of the cubic box.

We infer posterior parameter distributions for our model parameters by
performing Markov Chain Monte Carlo (MCMC) sampling using the parallel
affine-invariant ensemble sampler of Goodman and Weare
\cite{GoodmanWeare_2010} implemented in the {\sc{emcee}} python module
\cite{Foreman-Mackey_et_al_2016}.  In all cases we use 1000 walkers and
remove the first $10^5$ burn-in samples.

\begin{table}[H]
   \centering \caption{Parameter priors used in our likelihood analysis and
   input values of HOD parameters to our two ELUCID galaxy mocks, one built without
    galaxy assembly bias~(Mock A) and the other with strong galaxy assembly
    bias~(Mock B).}
    \begin{tabular}{lccc}
    \hline
        Parameter &  Prior & Mock A & Mock B\\
    \hline
        $\siglogM$ & $[0.01, 0.80]$ & $0.20$ & $0.20$\\
    $\log M_\mathrm{min}$ & $[11.0, 12.5]$ & $11.95$  & $11.95$\\
    $\log M_1$ & $[12.0, 13.5]$ & $13.15$ & $13.15$\\
    $\alpha$ & $[0.6, 2.0]$ & $1.0$ & $1.0$\\
    $\Acon$ &  $\mathcal{N} (0.86, 0.12^2)$ & $0.86$& $0.86$\\
    \hline
        $\Qcen$ & $[-2.0, 2.0]$ & $0.0$ & $-0.5$\\
        $\Qsat$ & $[-2.0, 2.0]$ & $0.0$ & $+0.5$\\
    \hline
  \end{tabular}
\label{table:priors&values}
\end{table}

We have found that the parameter $M_0$ has negligible effect on all of our
observables and is poorly constrained; we therefore fix $\log M_0{=}11.40$
in all cases. We also adopt a Gaussian prior on $\Acon$ centered at
$\Acon{=}0.86$ with standard deviation $0.12$. This prior is motivated by
the results of Zu et al.\ \cite{Zu_et_al_2014} who applied an HOD analysis
to the clustering and lensing data for the same SDSS-DR7 catalog that we
use. For all other parameters we adopt uniform priors listed in Table
\ref{table:priors&values}. In all such cases our priors are
non-informative. Although our galaxy assembly bias parameters can
technically assume values in the range $(-\infty, \infty)$ the priors we
impose upon them are non-informative and cover all physically plausible
values of the two parameters.\footnote{A value of $|Q_\mathrm{cen}|{=}2.0$,
for instance, would mean that the effective value of $M_\mathrm{min}$ could
vary by an entire order of magnitude at fixed mass based on large scale
overdensity.}

\section{Model Tests Using ELUCID Mock Galaxies}
\label{sec:mock}

\begin{table*}[htb!]
\renewcommand*{\arraystretch}{1.3}
   \centering \caption{Posterior constraints of model parameters from the
   ELUCID mock with no galaxy assembly bias. The
    column `Haloes' refers to the simulation used to compute the
    likelihood. Values quoted
    are posterior modes with $68\%$ confidence intervals. The value of $\chi^2 /
    \mathrm{d.o.f}$ is calculated from the mean of the posterior samples. Input values are listed in the last row.}
    \begin{adjustbox}{center}
    \begin{tabular}{llccccccccc}
    \hline
    Data vector & Haloes & $\siglogM$ & $\log M_\mathrm{min}$ & $\log M_1$ & $\alpha$ & $\Qcen$ & $\Qsat$ & $\Acon$ & $\chi^2 / \mathrm{d.o.f}$ \\
    \hline
    $\{w_p(r_p), n_g \}$ & ELUCID & $0.34^{+0.30}_{-0.23}$ & $12.00^{+0.12}_{-0.07}$& $13.19^{+0.17}_{-0.13}$ & $0.97^{+0.14}_{-0.11}$ & $-0.01^{+0.22}_{-0.30}$ & $-0.04^{+0.70}_{-0.58}$ & $0.85^{+0.12}_{-0.12}$ & $2.01/14$\\
    $\{w_p(r_p), n_g \}$ & Bolshoi & $0.38^{+0.29}_{-0.26}$ & $11.99^{+0.12}_{-0.09}$ & $13.09^{+0.21}_{-0.16}$ & $0.88^{+0.10}_{-0.10}$ & $-0.05^{+0.21}_{-0.40}$ & $-0.59^{+1.10}_{-0.81}$ & $0.85^{+0.12}_{-0.12}$ & $7.33/14$\\
    \hline
    $\{w_p(r_p), p(N_8^g), n_g \}$ & ELUCID  & $0.20^{+0.05}_{-0.05}$ &  $11.95^{+0.03}_{-0.02}$ & $13.14^{+0.05}_{-0.06}$ & $0.96^{+0.05}_{-0.06}$ & $0.00^{+0.06}_{-0.06}$ & $0.01^{+0.09}_{-0.09}$ & $0.81^{+0.09}_{-0.08}$ & $6.68/34$\\
    $\{w_p(r_p), p(N_8^g), n_g \}$ & Bolshoi  & $0.19^{+0.09}_{-0.09}$ & $11.89^{+0.03}_{-0.03}$ & $13.11^{+0.07}_{-0.08}$ & $0.99^{+0.07}_{-0.07}$ & $0.04^{+0.07}_{-0.06}$ & ${-}0.20^{+0.11}_{-0.12}$ & $0.83^{+0.12}_{-0.10}$ & $44.92/34$\\
    \hline
        $\{ w_{p} (r_p | \tilde{N}^{g,i}_8)\rvert_{i{=}1\mbox{-}5}, n_g \}$ & ELUCID & $0.17^{+0.09}_{-0.09}$ & $11.94^{+0.03}_{-0.03}$ & $13.14^{+0.06}_{-0.05}$ & $0.98^{+0.05}_{-0.05}$ & $-0.02^{+0.06}_{-0.05}$ & $0.03^{+0.12}_{-0.10}$ & $0.83^{+0.10}_{-0.10}$ & $11.96/94$\\
        $\{ w_{p} (r_p | \tilde{N}^{g,i}_8)\rvert_{i{=}1\mbox{-}5}, n_g \}$ & Bolshoi & $0.20^{+0.10}_{-0.11}$ & $11.94^{+0.04}_{-0.03}$ & $13.01^{+0.08}_{-0.08}$ & $0.89^{+0.04}_{-0.05}$ & $-0.15^{+0.06}_{-0.06}$ & $0.18^{+0.17}_{-0.15}$ & $0.79^{+0.10}_{-0.13}$ & $81.48/94$\\
    \hline
        $\{ w_{p} (r_p | \tilde{N}^{g,i}_8)\rvert_{i{=}1\mbox{-}5}, p(N_8^g), n_g \}$ & ELUCID & $0.20^{+0.04}_{-0.04}$  & $11.95^{+0.02}_{-0.02}$ & $13.15^{+0.04}_{-0.04}$ & $0.96^{+0.04}_{-0.04}$ & $0.00^{+0.05}_{-0.04}$ & $0.01^{+0.07}_{-0.07}$ & $0.80^{+0.07}_{-0.06}$ & $18.72/114$\\
        $\{ w_{p} (r_p | \tilde{N}^{g,i}_8)\rvert_{i{=}1\mbox{-}5}, p(N_8^g), n_g \}$ & Bolshoi & $0.19^{+0.05}_{-0.06}$ & $11.92^{+0.02}_{-0.02}$ & $13.09^{+0.04}_{-0.06}$ & $0.96^{+0.04}_{-0.05}$ & $0.02^{+0.05}_{-0.05}$ & $-0.06^{+0.10}_{-0.12}$ & $0.79^{+0.09}_{-0.10}$ &  $120.53/114$\\
    \hline
    Input Value & - & 0.20 & 11.95 & 13.15 & 1.0 & 0.0 & 0.0 & 0.86 & - \\
    \hline
  \end{tabular}
  \end{adjustbox}
\label{table:mock_constraints}
\end{table*}

\begin{table*}[htb!]
\renewcommand*{\arraystretch}{1.3}
    \centering \caption{Similar to Table~\ref{table:mock_constraints} but
    for the mock with significant level of galaxy assembly bias.  }
    \begin{adjustbox}{center}
    \begin{tabular}{llccccccccc}
    \hline
    Data vector & Haloes & $\siglogM$ & $\log M_\mathrm{min}$ & $\log M_1$ & $\alpha$ & $\Qcen$ & $\Qsat$ & $\Acon$ & $\chi^2 / \mathrm{d.o.f}$ \\
    \hline
    $\{w_p(r_p), n_g \}$ & ELUCID & $0.32^{+0.30}_{-0.22}$ & $12.00^{+0.12}_{-0.07}$  & $13.18^{+0.17}_{-0.14}$ & $0.97^{+0.11}_{-0.10}$ & $-0.58^{+0.25}_{-0.19}$ & $0.67^{+0.64}_{-0.55}$ & $0.86^{+0.12}_{-0.12}$ & $1.23/14$\\
    $\{w_p(r_p), n_g \}$ & Bolshoi & $0.45^{+0.26}_{-0.31}$ & $12.01^{+0.11}_{-0.10}$  & $13.05^{+0.18}_{-0.15}$ & $0.91^{+0.11}_{-0.10}$ & $-0.51^{+0.32}_{-0.40}$ & $-0.03^{+1.14}_{-0.82}$ & $0.86^{+0.12}_{-0.12}$ & $7.07/14$\\
    \hline
    $\{w_p(r_p), p(N_8^g), n_g \}$ & ELUCID & $0.20^{+0.07}_{-0.07}$ & $11.96^{+0.04}_{-0.03}$ & $13.14^{+0.07}_{-0.07}$ & $0.97^{+0.06}_{-0.06}$ & $-0.50^{+0.05}_{-0.05}$ & $0.50^{+0.08}_{-0.08}$ & $0.82^{+0.12}_{-0.11}$ & $3.53/34$\\
    $\{w_p(r_p), p(N_8^g), n_g \}$ & Bolshoi & $0.13^{+0.11}_{-0.08}$ & $11.90^{+0.05}_{-0.04}$ & $12.91^{+0.11}_{-0.10}$ & $0.88^{+0.08}_{-0.07}$ & $-0.26^{+0.08}_{-0.09}$ & $0.01^{+0.17}_{-0.13}$ & $0.85^{+0.12}_{-0.11}$ & $ 33.34/34$\\
    \hline
    $\{ w_{p} (r_p | \tilde{N}^{g,i}_8)\rvert_{i{=}1\mbox{-}5}, n_g \}$ & ELUCID & $0.17^{+0.10}_{-0.09}$ & $11.94^{+0.03}_{-0.03}$ & $13.12^{+0.09}_{-0.09}$ & $0.99^{+0.05}_{-0.05}$ & $-0.51^{+0.06}_{-0.06}$ & $0.50^{+0.18}_{-0.15}$ & $0.88^{+0.11}_{-0.11}$ & $11.53 / 94$ \\
    $\{ w_{p} (r_p | \tilde{N}^{g,i}_8)\rvert_{i{=}1\mbox{-}5}, n_g \}$ & Bolshoi & $0.20^{+0.09}_{-0.08}$ & $11.93^{+0.04}_{-0.03}$ & $13.01^{+0.07}_{-0.07}$ & $0.92^{+0.04}_{-0.04}$ & $-0.50^{+0.07}_{-0.08}$ & $0.41^{+0.21}_{-0.15}$ & $0.81^{+0.12}_{-0.12}$ & $ 72.73 / 94$ \\
    \hline
    $\{ w_{p} (r_p | \tilde{N}^{g,i}_8)\rvert_{i{=}1\mbox{-}5}, p(N_8^g), n_g \}$ & ELUCID & $0.20^{+0.04}_{-0.05}$ & $11.95^{+0.02}_{-0.02}$ & $13.15^{+0.04}_{-0.04}$ & $0.98^{+0.04}_{-0.04}$ & $-0.50^{+0.03}_{-0.04}$ & $0.50^{+0.05}_{-0.05}$ & $0.81^{+0.08}_{-0.08}$ & $14.71/114$\\
    $\{ w_{p} (r_p | \tilde{N}^{g,i}_8)\rvert_{i{=}1\mbox{-}5}, p(N_8^g), n_g \}$ & Bolshoi & $0.18^{+0.06}_{-0.07}$ & $11.91^{+0.03}_{-0.02}$ & $12.96^{+0.06}_{-0.06}$ & $0.92^{+0.04}_{-0.04}$ & $-0.37^{+0.06}_{-0.07}$ & $0.17^{+0.17}_{-0.11}$ & $0.76^{+0.12}_{-0.12}$ & $109.19/114$\\
    \hline
    Input Value & - & 0.20 & 11.95 & 13.15 & 1.0 & -0.50 & 0.50 & 0.86 & - \\
    \hline
  \end{tabular}
  \end{adjustbox}
\label{table:app_mock_constraints}
\end{table*}

\begin{figure*}[htb!]
\centering \includegraphics[width=1.0\textwidth]{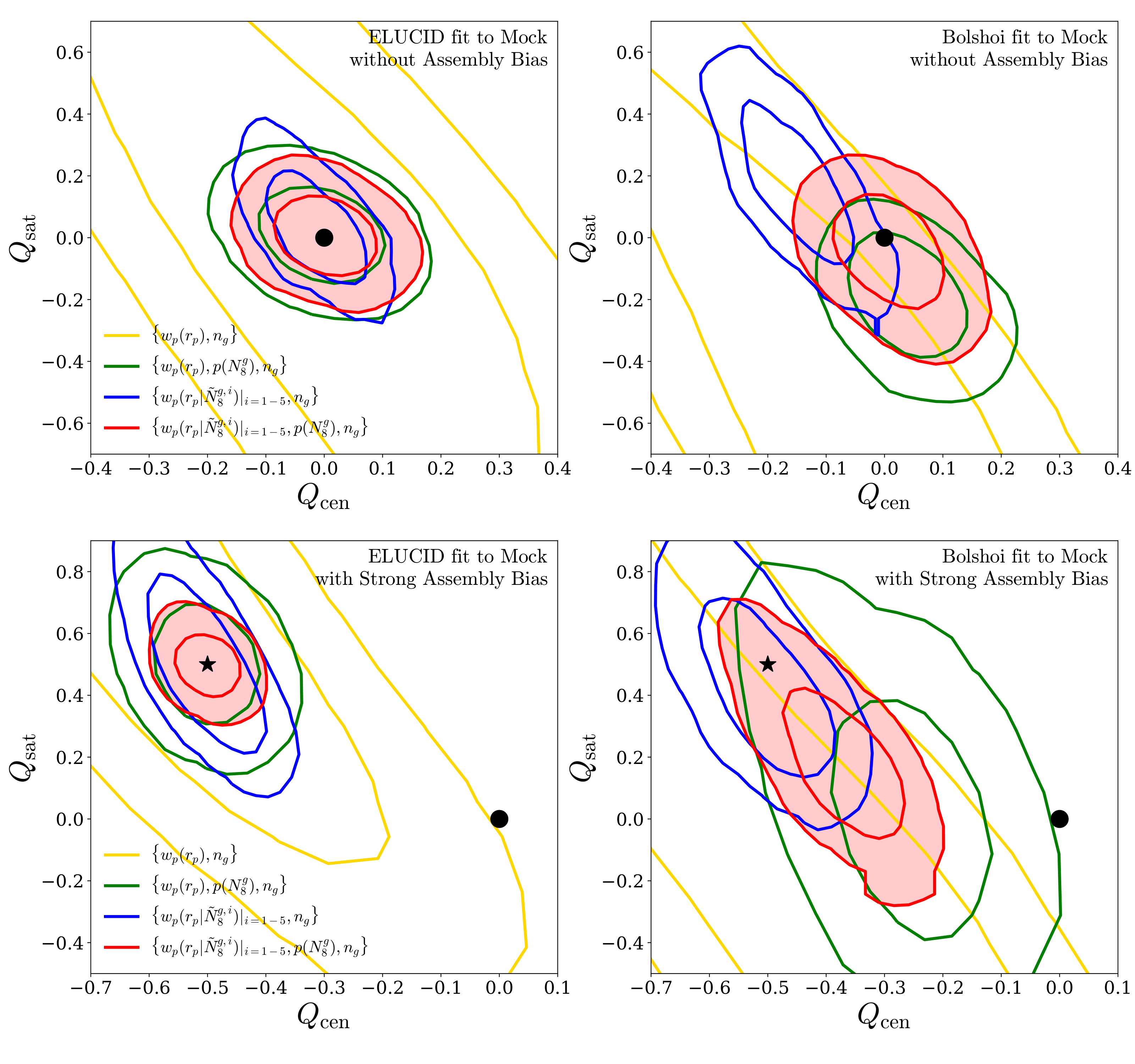}
\caption{Posterior constraints on $\Qcen$ and $\Qsat$ from fitting either
    ELUCID (left) or Bolshoi~(right) haloes to the ELUCID mock
    with~(bottom) or  without~(top) assembly bias. In each panel, we show
    the $68\%$ and $95\%$ confidence regions derived from using the
    $\{w_p(r_p), n_g\}$~(yellow), $\{w_p(r_p), p(\delta_8^g), n_g
    \}$~(green), $\{ w_{p} (r_p | R_\delta^{i})\rvert_{i{=}1\mbox{-}5}, n_g
    \}$~(blue), and $\{ w_{p} (r_p | R_\delta^{i})\rvert_{i{=}1\mbox{-}5},
    p(\delta_8^g), n_g \}$~(red) data vectors, respectively. The black
    circle marks $\Qcen{=}\Qsat{=}0.0$, while the black star marks the
    input values for populating the ELUCID galaxy mock with strong galaxy assembly
    bias. Red shading
    highlights our fiducial choice of data vector.}
\label{fig:2panel_mock}
\end{figure*}

Before analysing data from SDSS, we perform a variety of mock tests that
will be described below. In addition to allowing us to confirm the
robustness of our method, these mock tests compare the relative precision
of different data vectors, and investigate the level of systematic bias in
parameter estimation due to cosmic variance.  The values of the HOD
parameters used to populate the two mocks, one with zero and the other with
strong galaxy assembly bias, are listed in Table \ref{table:priors&values}.

We select the values of input HOD parameters for our mock with no galaxy
assembly bias~(Mock A) that produce reasonably good agreement with our
observables measured in SDSS, and only change the values of $\Qcen$ and
$\Qsat$ while keeping the other parameters fixed for our mock with strong
galaxy assembly bias~(Mock B).  In what follows we will refer to our HOD
analysis as ``fitting ELUCID or Bolshoi haloes to ELUCID or SDSS
galaxies''. By this we mean that we populate ELUCID or Bolshoi haloes with
our HOD prescription to compute model likelihoods to fit to data vectors
calculated using either ELUCID mock galaxies or SDSS galaxies.  For each of
the four data vectors, we perform four separate mock tests,
two by fitting ELUCID haloes to ELUCID mock galaxies constructed with and
without galaxy assembly bias, and the other two by fitting Bolshoi haloes to
the same ELUCID mock galaxies.  The results of our 16 mock constraints from
this section are listed in Table \ref{table:mock_constraints} using Mock A
and Table \ref{table:app_mock_constraints} using Mock B; values quoted correspond
to parameter posterior
modes and $68\%$ confidence intervals. The results of an analogous set of
tests for Mock B, but with the values of  $\Qcen$ and $\Qsat$ kept fixed to zero,
can be found in \ref{app:noGABtest}.

\subsubsection{The $\{w_p(r_p), n_g \}$ Data Vector}

All the standard HOD parameters listed in
Table~\ref{table:mock_constraints} and~\ref{table:app_mock_constraints} are
correctly recovered in all four cases,
albeit sometimes with large uncertainties (especially $\siglogM$) or
dominated by priors~(e.g., $\Acon$). In each case, the constraint is
statistics-limited, so that the cosmic variance between ELUCID and Bolshoi
is not discernible in either the goodness of fit or the size of the
uncertainties.  For all four cases, the marginalized 1D constraints on $\Qcen$
and $\Qsat$ are consistent with the input, but the very large uncertainties
render the constraint largely useless.

The marginalized 2D joint constraints in the $\Qcen$ vs. $\Qsat$ plane are
highlighted in Figure~\ref{fig:2panel_mock} as yellow contours in each
panel.  A strong degeneracy of anti-correlation
between the two parameters emerges in the 2D constraints, and in the
Bolshoi case the degeneracy drives a ${\sim}1.5\sigma$ deviation from the
true values of $\Qsat$ and $\Qcen$~(black circle and star in the top and
bottom panels, respectively).  The degeneracy track is unsurprising, as a
change in $\Qcen$ can always be partially compensated by a change of
$\Qsat$ of the opposite sign.  The deviation, however, is either a result
of the slight difference in cosmology or of cosmic variance.  We consider
it unlikely that this difference is caused by cosmology for two reasons.
Firstly, the actual difference in cosmological parameters between the two
simulations is reasonably small~(e.g., $\sigma_8{=}0.80$ vs. $0.82$)
relative to the expected sensitivity of $w_p$ on these
parameters\footnote{See, for example, figure 4 of
\cite{Salcedo_et_al_2020}.}.  Secondly, if the discrepancy were
attributable to cosmology we would expect that our other HOD parameters
would be similarly affected if not more so. Unlike our four traditional HOD
parameters, both $\Qcen$ and $\Qsat$ act to decouple the one and two-halo
scales, and they also introduce a unique scale dependence to the
correlation function. Therefore, we believe that any discrepancy between
the ELUCID and Bolshoi constraints is primarily driven by the cosmic
variance as observed in Figure~\ref{fig:footprint}.

Overall, we find that $\{ w_p(r_p), n_g\}$ is not very effective at
constraining galaxy assembly bias for an SDSS-like sample, but it provides
the baseline constraints that we will compare to when examining the
constraints from more advanced data vectors below.

\subsubsection{The $\{w_p(r_p), p(N_8^g), n_g \}$ Data Vector}

We now turn to mock tests performed with the $\{w_p(r_p), p(N_8^g), n_g \}$
data vector. Compared to the $\{ w_p(r_p), n_g \}$ tests, adding $p(N_8^g)$
produces drastically improved the precision of the constraints. Our constraint on $\siglogM$ has
sharpened to $25\%$, an improvement of a factor of five.  Similarly the
constraints on the parameters $\log M_\mathrm{min}$ and $\log M_1$ have
sharpened considerably by a factor of about three, while the constraint on
$\alpha$ has also improved by a factor of 2-3.

Focusing on our galaxy assembly bias parameters $\Qcen$ and $\Qsat$, Figure
\ref{fig:2panel_mock} shows their constraints in green contours in each
panel.  The $\{w_p(r_p), p(N_8^g), n_g\}$ data vector also significantly
improves the precision of the constraints in all cases, but the inclusion
of $p(N_8^g)$ tends to shift the constraints towards the high-$\Qcen$,
low-$\Qsat$ direction in the Bolshoi fits~(right panels). For Mock A with
zero galaxy assembly bias, the 2D joint constraint is
still consistent with the input at ${\sim}1\sigma$~(top right panel), but the
shift produces a much stronger bias~(${>}2\sigma$) for Mock B when the input is far from
$\Qsat=\Qcen=0$~(bottom right panel). In the two left panels when ELUCID
haloes are used, the $\{w_p(r_p), p(N_8^g), n_g\}$ data vector
yields constraints that are not only tightened but also stay unbiased.

Therefore, despite the difference of detail between $p(N_8^g)$ and the
count-in-cell statistics used by Wang et al.\ \cite{Wang_et_al_2019}, we
confirm their finding that by combining one and two-point statistics one
can obtain an improved constraint on galaxy assembly bias, but
with the caveat that the impact of cosmic variance is greatly mitigated.

\subsubsection{The $\{ w_{p} (r_p | \tilde{N}^{g,i}_8)\rvert_{i{=}1\mbox{-}5}, n_g \}$ Data Vector}

Next we examine mock tests performed with the $\{ w_{p} (r_p |
\tilde{N}^{g,i}_{8})\rvert_{i{=}1\mbox{-}5}, n_g \}$ data vector.
Overall, the 1D constraints on the standard HOD parameters are very similar
compared to those using $\{w_p(r_p), p(N_8^g), n_g \}$, suggesting that the
constraining power on the global HOD is comparable between
$w_p(r_p)+p(N_8^g)$ and $w_{p} (r_p |
\tilde{N}^{g,i}_8)\rvert_{i{=}1\mbox{-}5}$. This is unsurprising because
the combination of five count-dependent correlation functions should
contain all the information encoded in the global $w_p(r_p)$ and some of
the key information on the shape of the galaxy number count distribution.

In the left panels of Figure~\ref{fig:2panel_mock}), the constraints
derived using $\{w_{p} (r_p | \tilde{N}^{g,i}_8)\rvert_{i{=}1\mbox{-}5},
n_g\}$ (blue contours) also exhibit similar levels of accuracy compared to those
using $\{w_p(r_p), p(N_8^g), n_g \}$ (green contours), with the input
values of $\Qcen$ and $\Qsat$ correctly recovered regardless of having
zero~(top left) or strong~(bottom left) galaxy assembly bias. When the
Bolshoi haloes are used~(right panels), however, the constraints using
$w_{p} (r_p | \tilde{N}^{g,i}_8)\rvert_{i{=}1\mbox{-}5}, n_g $ shift to the
low-$\Qcen$, high-$\Qsat$ regime, in the opposite direction compared to the
shift caused by including $p(N_8^g)$~(green contours). This shift produces
a ${\sim}2\sigma$ bias in the $\Qcen$-$\Qsat$ plane when there is no galaxy
assembly bias~(top right), but stays within $1\sigma$ when the galaxy
assembly bias is strong~(bottom right).

Combining the test results from using $\{w_p(r_p), p(N_8^g), n_g \}$ (green
contours) and $\{w_{p} (r_p | \tilde{N}^{g,i}_8)\rvert_{i{=}1\mbox{-}5},
n_g\} $ (blue contours), we find that although each data vector can
produce biased constraints at the $2\sigma$ level in the presence of strong
cosmic variance, $p(N_8^g)$ and $w_{p} (r_p |
\tilde{N}^{g,i}_8)\rvert_{i{=}1\mbox{-}5}$ shift the constraint in the
opposite direction with each other. This behavior indicates that the galaxy
assembly bias information present in the two observables are relatively
independent, and by combining the two~(i.e. our fiducial data vector below)
we can potentially break some of the degeneracy between galaxy assembly
bias and cosmic variance.

\begin{figure*}[htb!]
\centering \includegraphics[width=1.0\textwidth]{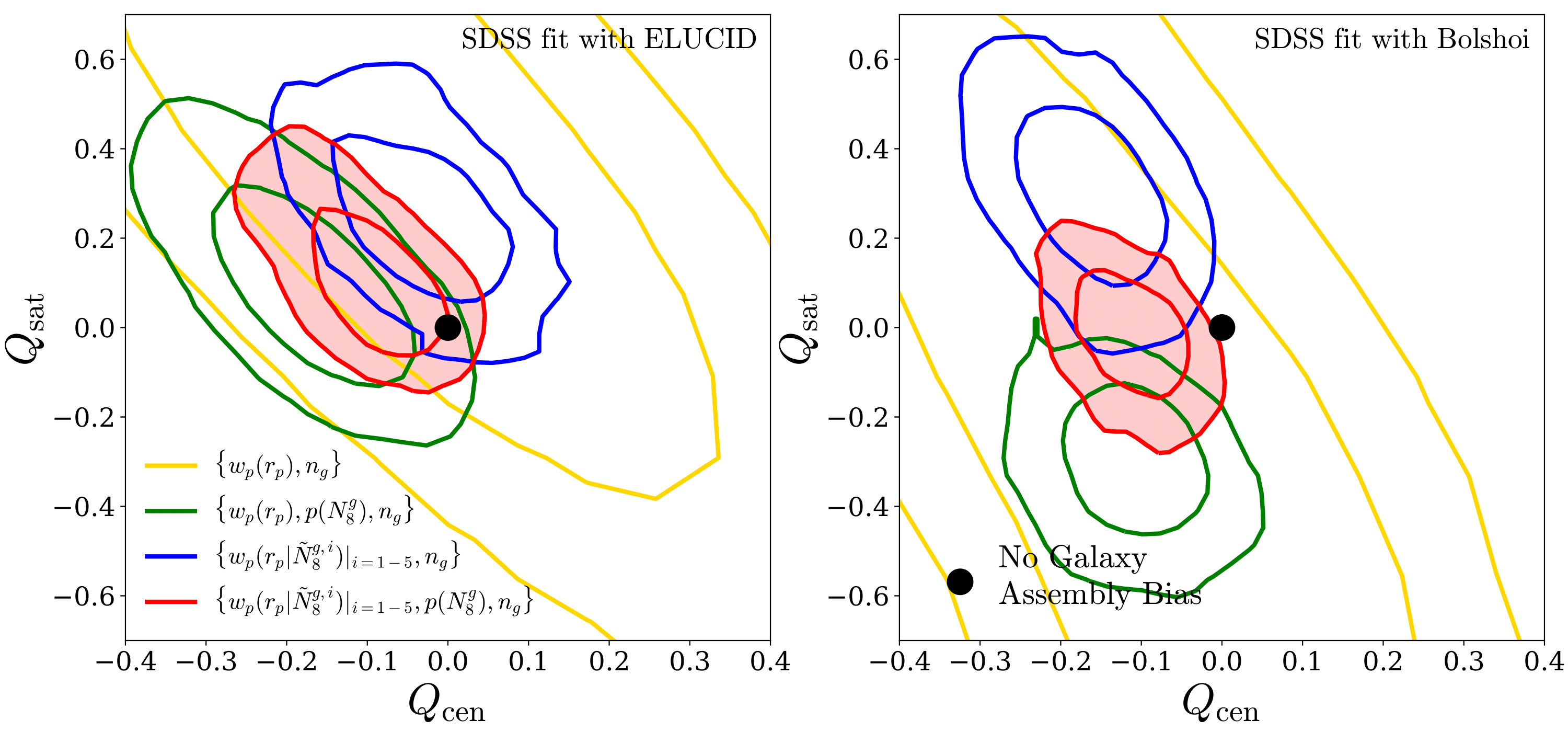}
\caption{Similar to Figure~\ref{fig:2panel_mock} but for constraints from
the SDSS galaxies, with black circles indicating  $\Qcen{=}\Qsat{=}0.0$ as
a visual reference to the case of no assembly bias. Red shading highlights
the constraint from our fiducial choice of data vector.}
\label{fig:2panel_mstardata}
\end{figure*}

\subsubsection{The $\{ w_{p} (r_p | \tilde{N}^{g,i}_8)\rvert_{i{=}1\mbox{-}5}, p(N_8^g), n_g \}$ Data Vector}

\begin{figure*}[htb!]
\centering \includegraphics[width=1.0\textwidth]{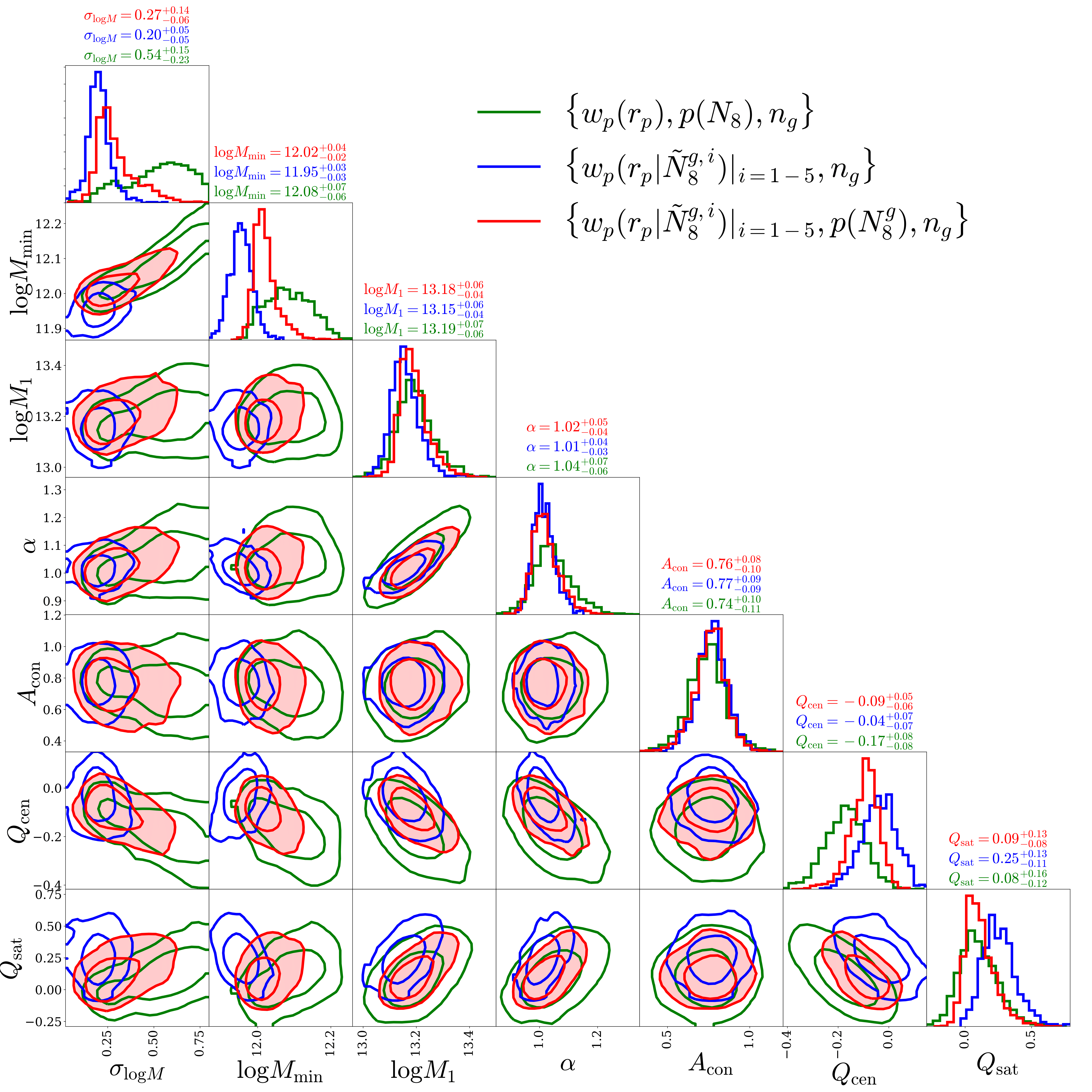}
\caption{Posterior constraints from our MCMC analysis of the SDSS data,
derived from $\{w_p(r_p), p(N_8^g), n_g\}$~(green), $\{ w_{p} (r_p |
    \tilde{N}_8^{g,i})\rvert_{i{=}1\mbox{-}5}, n_g\}$~(blue), and our
    fiducial data vector~$\{w_{p} (r_p |
    \tilde{N}_8^{g,i})\rvert_{i{=}1\mbox{-}5}, p(N_8^g), n_g \}$~(red
    filled contours and red histograms). Contours in each off-diagonal
    panel show the $68\%$ and $95\%$ confidence regions, and histograms in
    each diagonal panel show the 1D marginalized posterior distribution of
    each parameter, with values quoted above
    corresponding to the posterior mode and $68\%$ confidence intervals.}
\label{fig:mstardata_fullcorner}
\end{figure*}

Finally, we examine the mock tests performed with $\{ w_{p} (r_p |
\tilde{N}^{g,i}_8)\rvert_{i{=}1\mbox{-}5}, p(N_8^g), n_g \}$, the input
data vector that will be adopted for our fiducial constraint when analysing
the SDSS data.  In general, the constraints on the standard HOD parameters
are marginally improved compared to the $\{ w_{p} (r_p |
\tilde{\delta}^{g,i}_8)\rvert_{i{=}1\mbox{-}5}, n_g \}$ case, except for
$\siglogM$, for which the constraint improves by about a factor of two.

More important, this data vector provides more robust constraints on the
galaxy assembly bias parameters. For the ELUCID fit to Mock A~(top left),
the derived constraint is $\Qcen{=}{-}0.00^{+0.05}_{-0.04}$ and
$\Qsat{=}0.01^{+0.07}_{-0.07}$, in excellent agreement with the input
values. For the Bolshoi fit to Mock A~(top right), adding $p(N_8^g)$ brings
the constraints significantly closer to the input values compared to the
large discrepancy observed when only $\{ w_{p} (r_p |
\tilde{N}^{g,i}_8)\rvert_{i{=}1\mbox{-}5}, n_g \}$ is used, from
$\Qcen{=}{-}0.15^{+0.06}_{-0.06}$ and $\Qsat{=}0.18^{+0.17}_{-0.15}$, a
$2\sigma$ deviation in this 2D space, to $\Qcen{=}0.02^{+0.05}_{-0.05}$ and
$\Qsat{=}{-}0.06^{+0.10}_{-0.12}$~(i.e., within $1\sigma$ deviation). This
improvement is reassuring --- the residual cosmic variance between ELUCID
and SDSS should be much smaller than that between Bolshoi and ELUCID, so we
expect the amount of bias in our constraints on $\Qcen$ and $\Qsat$ to be
significantly smaller than the statistical uncertainties. Meanwhile, the
fits to Mock B exhibit similar results, except that the deviation increases
in the Bolshoi fit~(bottom right) but still within $2\sigma$.

\begin{figure*}[htb!]
\centering \includegraphics[width=1.0\textwidth]{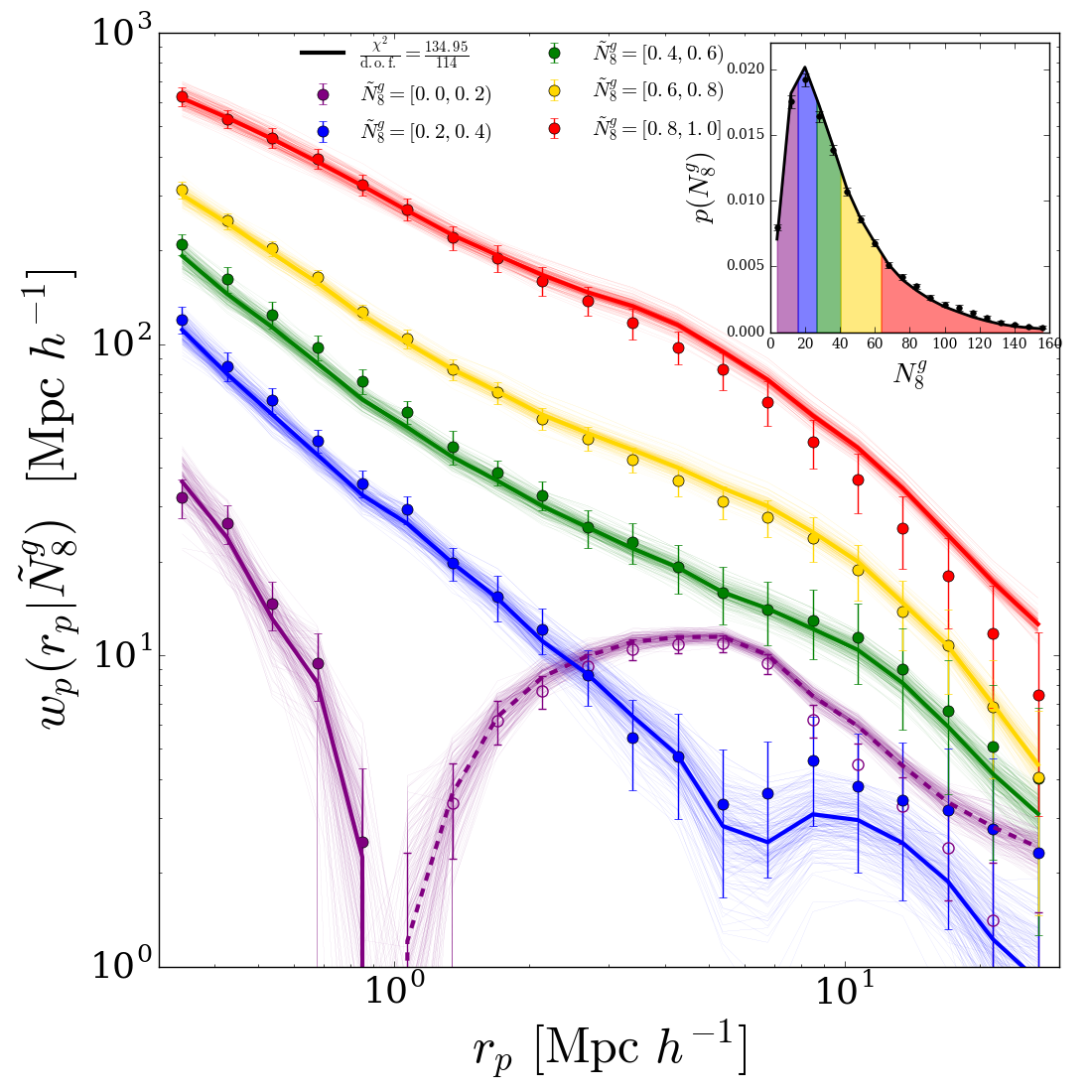}
\caption{Comparison of the measured projected cross-correlation functions
between galaxy quintiles selected in $N_8^g$ and the
    global galaxy sample~(circles with errorbars) and those predicted by
    our posterior mean model with ELUCID~(thick solid curves). Thin bundle
    of light-coloured curves surrounding each thick solid curve represent
    the predictions from 100 random steps along the MCMC chain of our SDSS
    analysis.  The inset panel shows the data~(black circles with
    errorbars) vs. model prediction~(solid curve) comparison for
    $p(\del8)$, with the five filled vertical bands underneath the black
    curve indicating the quintiles in $N_8^g$. Note that the legend gives
    values of the rank quantity $\tilde{N}_8^g$ not $N_8^g$.  The posterior
    model produces a good description to the data with a $\chi^2$
    value of $134.95$ for $114$ degrees of freedom.}
\label{fig:best_fit}
\end{figure*}

To summarize, through a variety of mock tests with ELUCID and Bolshoi
haloes, we find that our galaxy assembly bias constraints from the
$\{w_p(r_p), n_g\}$, $\{w_p(r_p), p(N_8^g), n_g\}$ and, $\{w_{p} (r_p |
\tilde{N}^{g,i}_8)\rvert_{i{=}1\mbox{-}5}, n_g \}$ data vectors are
sensitive to the impact of cosmic variance. In contrast, the constraints
from the $\{ w_{p} (r_p |
\tilde{N}^{g,i}_8)\rvert_{i{=}1\mbox{-}5},p(N_8^g), n_g \}$ data vector are
largely unbiased, despite the strong level of cosmic variance between
ELUCID and Bolshoi. Among the four data vectors, $\{ w_{p} (r_p |
\tilde{N}^{g,i}_8)\rvert_{i{=}1\mbox{-}5}, p(N_8^g), n_g \}$ also produces
the most stringent constraints on the galaxy assembly bias parameters.
Therefore, for our analysis of the real data, we will apply our extended
HOD model to the ELUCID simulation to fit the $\{ w_{p} (r_p |
\tilde{N}^{g,i}_8)\rvert_{i{=}1\mbox{-}5}, p(N_8^g), n_g \}$ data vector
measured from SDSS as our fiducial analysis.

\section{Galaxy Assembly Bias Constraint from SDSS}
\label{sec:result}

\begin{table*}[htb!]
\renewcommand*{\arraystretch}{1.3}
   \centering
    \caption{Posterior constraints of model parameters from the SDSS data. The column `Haloes' refers to the simulation used to compute the likelihood. Values quoted are posterior modes with $68\%$ confidence intervals. The value of $\chi^2 / \mathrm{d.o.f}$ is calculated from the mean of the posterior samples.}
    \begin{adjustbox}{center}
    \begin{tabular}{llcccccccc}
    \hline
    Data vector & Haloes & $\siglogM$ & $\log M_\mathrm{min}$ & $\log M_1$ & $\alpha$ & $\Qcen$ & $\Qsat$ & $\Acon$ & $\chi^2 / \mathrm{d.o.f}$ \\
    \hline
    $\{w_p(r_p), n_g \}$ & ELUCID & $0.39^{+0.24}_{-0.22}$ & $11.99^{+0.11}_{-0.06}$ & $13.28^{+0.12}_{-0.10}$ & $1.09^{+0.09}_{-0.09}$ & $-0.06^{+0.26}_{-0.22}$ & $0.40^{+0.49}_{-0.46}$ & $0.81^{+0.12}_{-0.13}$ & $14.23/14$ \\
    $\{w_p(r_p), n_g \}$ & Bolshoi& $0.30^{+0.28}_{-0.20}$ & $11.92^{+0.11}_{-0.06}$ & $13.24^{+0.15}_{-0.11}$ & $1.10^{+0.10}_{-0.08}$ & $-0.08^{+0.18}_{-0.26}$ & $-0.25^{+0.71}_{-0.59}$ & $0.85^{+0.12}_{-0.13}$ & $13.75/14$ \\
    \hline
    $\{w_p(r_p), p(N_8^g), n_g \}$ & ELUCID & $0.54^{+0.15}_{-0.23}$ & $12.08^{+0.07}_{-0.06}$ & $13.19^{+0.07}_{-0.06}$ & $1.04^{+0.07}_{-0.06}$ & $-0.17^{+0.08}_{-0.08}$ & $0.08^{+0.16}_{-0.12}$ & $0.74^{+0.10}_{-0.11}$ & $46.48/34$ \\
    $\{w_p(r_p), p(N_8^g), n_g \}$ & Bolshoi & $0.22^{+0.09}_{-0.07}$ & $11.95^{+0.03}_{-0.02}$ & $13.13^{+0.06}_{-0.05}$ & $1.02^{+0.06}_{-0.05}$ & $-0.11^{+0.06}_{-0.06}$ & $-0.29^{+0.11}_{-0.11}$ & $0.78^{+0.10}_{-0.09}$ & $41.13/34$ \\
    \hline
        $\{ w_{p} (r_p | \tilde{N}^{g,i}_8)\rvert_{i{=}1\mbox{-}5}, n_g \}$ & ELUCID & $0.20^{+0.05}_{-0.05}$ & $11.95^{+0.03}_{-0.03}$ & $13.15^{+0.06}_{-0.04}$ & $1.01^{+0.04}_{-0.03}$ & $-0.04^{+0.07}_{-0.07}$ & $0.25^{+0.13}_{-0.11}$ & $0.77^{+0.09}_{-0.09}$ & $76.28/94$ \\
        $\{ w_{p} (r_p | \tilde{N}^{g,i}_8)\rvert_{i{=}1\mbox{-}5}, n_g \}$ & Bolshoi & $0.19^{+0.06}_{-0.06}$ & $11.97^{+0.03}_{-0.03}$ & $13.07^{+0.05}_{-0.07}$ & $0.93^{+0.03}_{-0.04}$ & $-0.16^{+0.06}_{-0.06}$ & $0.29^{+0.14}_{-0.12}$ & $0.74^{+0.09}_{-0.11}$ &  $152.31/94$ \\
    \hline
        $\{ w_{p} (r_p | \tilde{N}^{g,i}_8)\rvert_{i{=}1\mbox{-}5}, p(N_8^g), n_g \}$ & ELUCID & $0.27^{+0.14}_{-0.06}$ & $12.02^{+0.04}_{-0.02}$ & $13.18^{+0.06}_{-0.04}$ & $1.02^{+0.05}_{-0.04}$ & $-0.09^{+0.05}_{-0.06}$ & $0.09^{+0.13}_{-0.08}$ & $0.76^{+0.08}_{-0.10}$ & $125.11 /114$\\
        $\{ w_{p} (r_p | \tilde{N}^{g,i}_8)\rvert_{i{=}1\mbox{-}5}, p(N_8^g), n_g \}$ & Bolshoi & $0.20^{+0.05}_{-0.04}$ & $11.98^{+0.02}_{-0.02}$ & $13.12^{+0.04}_{-0.04}$ & $0.99^{+0.03}_{-0.03}$ & $-0.11^{+0.04}_{-0.05}$ & $-0.02^{+0.09}_{-0.08}$ & $0.77^{+0.07}_{-0.09}$ & $167.22/114$ \\
    \hline
  \end{tabular}
  \end{adjustbox}
\label{table:data_constraints}
\end{table*}

Informed by the extensive mock tests conducted in \S~\ref{sec:mock}, we
conclude that the $\{ w_{p} (r_p |
\tilde{N}^{g,i}_8)\rvert_{i{=}1\mbox{-}5}, p(N_8^g), n_g \}$ data vector is
the optimal choice for the input measurements to our fiducial constraint.
We nonetheless fit ELUCID and Bolshoi haloes to the SDSS data using all
four data vectors listed in \S~\ref{sec:model}. We summarize the results of
the 1D constraints in Table \ref{table:data_constraints} and show the 2D
constraints for all the parameter pairs in
Figure~\ref{fig:mstardata_fullcorner}, respectively.  Overall, the
constraints derived from the SDSS galaxies are very similar to those in the
mock tests from the ELUCID mock galaxies, in terms of both the precision
using each data vector and the trends with data vectors/simulations.

The constraints on the galaxy assembly bias parameters are highlighted in
the left~(ELUCID) and right~(Bolshoi) panels of Figure
\ref{fig:2panel_mstardata}, with the red filled contours indicating the
results from our fiducial data vector $\{ w_{p} (r_p |
\tilde{N}_8^{g,i})\rvert_{i{=}1\mbox{-}5}, p(N_8^g), n_g \}$.  In each
panel, yellow, green, and blue open contours represent the constraints from
using $\{w_{p}(r_p), n_g\}$, $\{w_{p}(r_p), p(N_8^g), n_g\}$, and $\{ w_{p}
(r_p |\tilde{N}_8^{g,i})\rvert_{i{=}1\mbox{-}5}, n_g \}$, respectively.
Again, we observe similar behavior of each data vector using either
simulation as in Figure~\ref{fig:2panel_mstardata}. In particular, the
$\{w_{p}(r_p), n_g\}$ constraints are the weakest with strong degeneracy
exhibited between $\Qcen$ and $\Qsat$. Adding $p(N_8^g)$ tightens the
constraints significantly while bringing the contours into ${\sim}1.5\sigma$
tension with the zero values of $\Qcen$ and $\Qsat$. Swapping $w_{p} (r_p |
\tilde{\delta}_8^{g,i})\rvert_{i{=}1\mbox{-}5}$ for $p(N_8^g)$ further
tightens the constraints, and exhibits similar deviation from
$\Qcen{=}\Qsat{=}0$ --- the blue contours are inconsistent with zero galaxy
assembly bias at the $1.5\sigma$~(ELUCID) and $>2\sigma$~(Bolshoi) levels,
respectively. However, because of the analogous but well-understood
behaviors shown in our high-fidelity mock tests, we can safely attribute
the discrepancy largely to the (residual) cosmic variance between the
simulations and SDSS.  Finally, the fiducial ELUCID constraints with $\{
w_{p,gg} (r_p | \tilde{N}_8^{g,i})\rvert_{i{=}1\mbox{-}5}, p(N_8^g),
n_g\}$~(red filled contours on the left panel) are consistent with
$\Qcen{\equiv}\Qsat{=}0$ at $1\sigma$. Analogous constraints using
Bolshoi~(red filled contours on the right panel) are in $2\sigma$ tension
with $\Qcen{=}\Qsat{=}0$ a difference attributable to the large cosmic
variance between Bolshoi and SDSS.

Therefore, based on our fiducial SDSS analysis using the constrained
simulation ELUCID, we obtain a stringent marginalized 1D constraint of
$\Qcen{=}-0.09^{+0.05}_{-0.06}$ and $\Qsat{=}0.09^{+0.13}_{-0.08}$, hinting
at a significant detection of galaxy assembly bias at above $1\sigma$ level
and close to $2\sigma$~(at least for the centrals).  However, the
marginalized 2D constraint on the $\Qcen$-$\Qsat$ plane~(red contours on
the left panel of Figure \ref{fig:2panel_mstardata}) is consistent with
having no assembly bias at $1\sigma$ and well within $2\sigma$. Given that there is still some
residual impact of cosmic variance between ELUCID and SDSS that is not
included in our uncertainties, our current fiducial constraint indicates no
evidence for having a strong halo assembly bias in the local Universe for
the population of galaxies selected by
$M_*{\geq}10^{10.2}\,h^{-2}\,M_\odot$, typical of galaxies in the SDSS Main
Galaxy redshift survey. In Figure \ref{fig:best_fit} we show the comparison
between the overdensity-dependent cross-correlation functions predicted by
our best-fitting model with ELUCID~(coloured curves) and the data~(circles
with errorbars).  The inset panel shows the comparison for the PDF of
galaxy overdensity.  The best-fitting prediction yields a value of $\chi^2
/ \mathrm{d.o.f.} {=} 134.95/114$, indicating a satisfactory goodness of
fit to the data, with each of the observables well described at all scales.
This is an impressive testament to the ability of the combination of our
HOD model and the ELUCID reconstruction to describe the distribution and
clustering of galaxies in the local Universe, without the need to invoke
any strong galaxy assembly bias effect. Previous successes of HOD models in
reproducing the global $w_p(r_p)$ and 1-point PDF did not guarantee success
in reproducing the observed correlation functions of galaxies in $N_g^8$
quintiles, which provide a much more detailed characterization of
environment-dependent clustering.

\section{Conclusions}
\label{sec:conc}

In this paper, we have investigated the level of galaxy assembly bias in
the local Universe by performing a comprehensive HOD modelling of the
overdensity environment and projected clustering of SDSS galaxies, using
the state-of-the-art constrained simulation ELUCID that accurately
reconstructed the initial density perturbations of the SDSS volume.

For modelling galaxy assembly bias, we extend the standard HOD prescription
by including separate levels of central and satellite assembly bias,
parametrized by $\Qcen$ and $\Qsat$, respectively. In particular, we extend
the galaxy assembly bias parametrization of Wibking et al.\
\cite{Wibking_et_al_2019} and Salcedo et al.\ \cite{Salcedo_et_al_2020} to
include satellite assembly bias in the form of the parameter $\Qsat$ (see
also \cite{Xu_Zheng_2020}). The parameter $\Qsat$ allows the satellite
occupation to vary at fixed mass based on the large scale environment by
varying $\log M_1$ on a halo-by-halo basis. The parameter $\Qcen$ (called
$\mathcal{Q}_\mathrm{env}$ in \cite{Wibking_et_al_2019} and
\cite{Salcedo_et_al_2020}) acts similarly by modifying $\log
M_\mathrm{min}$.

We have employed a variety of data vectors consisting of one and two-point
galaxy statistics.  Extensive mock tests demonstrate that using a
non-constrained simulation could potentially show false evidence of galaxy
assembly bias for some of the data vectors that are sensitive to the cosmic
variance. Among the four data vectors we tested, we identify the
combination of correlation functions of galaxy quintiles selected by galaxy
number count $N_8^g$, the probability density distribution of $N_8^g$, and
the overall galaxy number density $n_g$, as the fiducial input for our
analysis with the SDSS data, because of its stringent constraining power.

Applying our extended HOD model on the ELUCID simulation to fit the SDSS
galaxies, our fiducial analysis yields stringent constraints of the level
of galaxy assembly bias in SDSS, with $\Qcen{=}-0.09^{+0.05}_{-0.06}$ and
$\Qsat{=}0.09^{+0.13}_{-0.08}$, respectively. Therefore, we do not find
evidence for the existence of a strong~($>2\sigma$) galaxy assembly bias within the SDSS main
galaxy sample, despite examining statistics that would be sensitive to such
bias if it were present.  The best-fitting model provides an excellent
description of the overdensity environment of each {\it individual}
observed galaxy, as well as the projected clustering of galaxies within
different overdensity environments, ranging from voids to average field to
clusters and super-clusters.

Although the combined one- and two-point statistics measured from SDSS are
consistent with having no galaxy assembly bias, our constraint is dominated
by the statistical uncertainties, with residual cosmic variance effect due
to imperfect ELUCID reconstruction in the underdense regions. With current
and upcoming spectroscopic surveys like Dark Energy Spectroscopic
Instrument (DESI; \cite{DESI_et_al_2016}) and the Prime Focus Spectrograph
(PFS; \cite{Takada_PFS_et_al_2014}), the uncertainties of our method will
be greatly reduced due to the orders-of-magnitude increase in the observed
number of spectra, as well as a thorough removal of the cosmic variance
with the ever-increasing survey volume.

\Acknowledgements{We thank the anonymous referees for helpful suggestions
that have greatly improved the manuscript, and Ben Wibking, Hao-Yi Wu, and
Zheng Zheng for stimulating discussions about this work. YZ and HYW
acknowledge the support by the National Key Basic Research and Development
Program of China (No.  2018YFA0404504). YZ acknowledges the support by the
National Science Foundation of China (11873038, 11621303, 11890692,
12173024), the science research grants from the China Manned Space Project
(No. CMS-CSST-2021-A01, CMS-CSST-2021-A02, CMS-CSST-2021-B01), the National
One-Thousand Youth Talent Program of China, and the STJU start-up fund (No.
WF220407220).  ANS thanks SJTU and TDLI for hospitality. ANS was supported
by a Department of Energy Computational Science Graduate Fellowship. This
material is based upon work supported by the U.S. Department of Energy,
Office of Science, Office of Advanced Scientific Computing Research,
Department of Energy Computational Science Graduate Fellowship under Award
Number DE-FG02-97ER25308. YZ, XHY, and YPJ acknowledge the support by the
``111'' project of the Ministry of Education under grant No. B20019.  YZ
thanks Cathy Huang for her hospitality during his visit to the Zhangjiang
Hi-Technology Park where he worked on this project. HYW is supported by the
National Science Foundation of China (11733004, 11421303, 11890693).  DW
acknowledges support of NSF grant AST 2009735. The Bolshoi simulations have
been performed within the Bolshoi project of the University of California
High-Performance AstroComputing Center (UC-HiPACC) and were run at the NASA
Ames Research Center. The CosmoSim database used in this paper is a service
by the Leibniz-Institute for Astrophysics Potsdam (AIP). Simulations were
analyzed in part on computational resources of the Ohio Supercomputer
Center \cite{OhioSupercomputerCenter1987}, with resources supported in part
by the Center for Cosmology and AstroParticle Physics at the Ohio State
University. We gratefully acknowledge the use of the {\sc{matplotlib}}
software package \cite{Hunter_2007} and the GNU Scientific library
\cite{GSL_2009}.  This research has made use of the SAO/NASA Astrophysics
Data System. This report was prepared as an account of work sponsored by an
agency of the United States Government. Neither the United State Government
nor any agency thereof, nor any of their employees, makes any warranty,
express or implied, or assumes any legal liability or responsibility for
the accuracy, completeness, or usefulness of any information, apparatus,
product, or process disclosed, or represents that its use would not
infringe on privately owned rights. Reference herein to any specific
commercial product, process, or service by trade name, trademark,
manufacturer or otherwise does not necessarily constitute or imply its
endorsement, recommendation, or favoring  by the United States Government
or any agency thereof.  The views and opinions of authors expressed herein
do not necessarily state or reflect those of the United State Government or
any agency thereof.}

\InterestConflict{The authors declare that they have no conflict of interest.}

\bibliographystyle{abbrv}
\bibliography{elucidbib_v2}

\begin{appendix}

\section{Covariance Matrices}
\label{app:cov}

\begin{figure}[H]
\centering \includegraphics[width=0.45\textwidth]{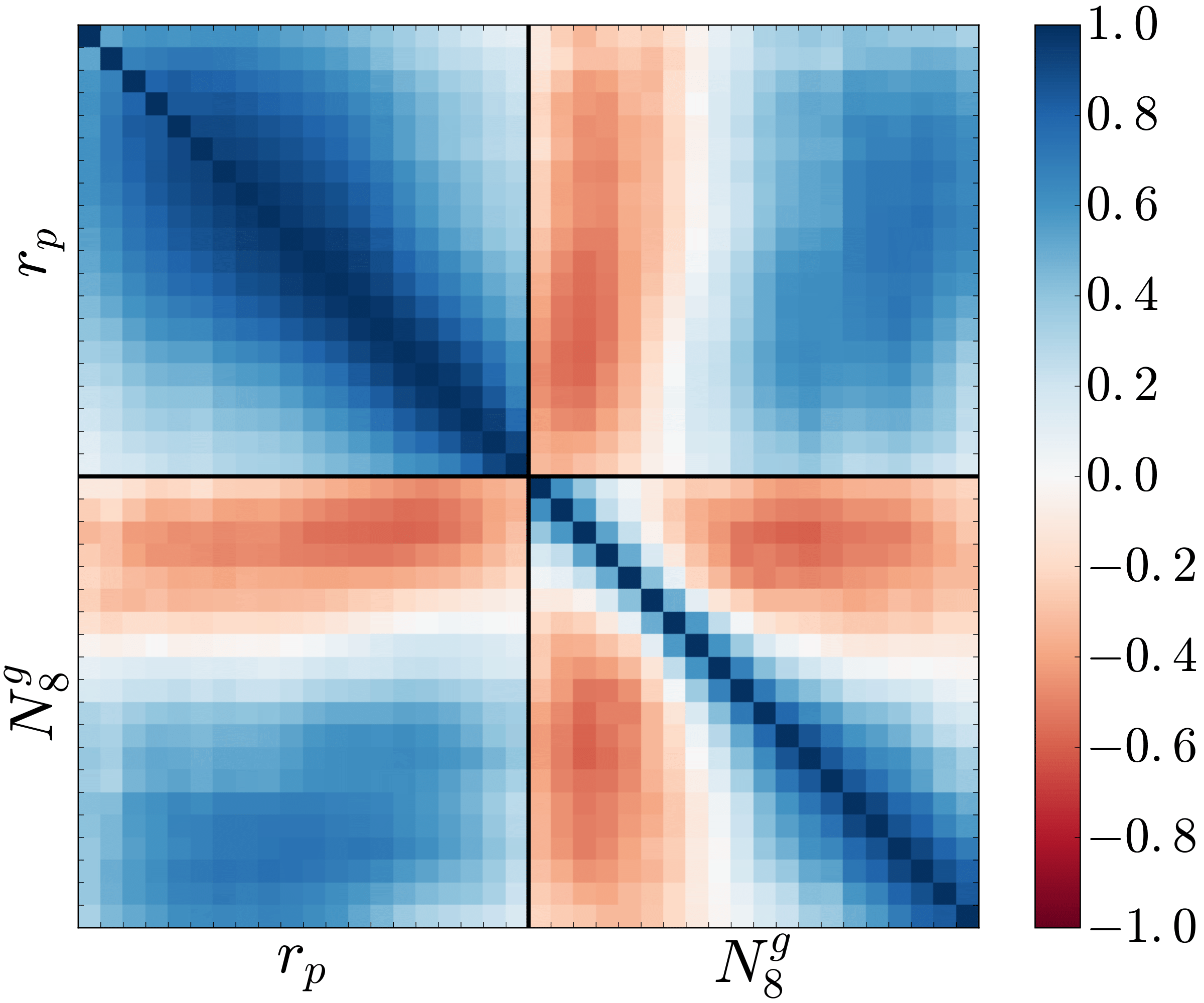}
\caption{Correlation matrix of $\{ w_{p}, p(N_8^g) \}$ calculated directly
    from SDSS data via jackknife re-sampling, with the correlation
    coefficients colour-coded by the colour bar on the right. Bins from
    left (top) to right (bottom) correspond to 20 logarithmically-spaced bins of
    $w_p(r_p)$ from $r_p{=}0.3\,h^{-1}\,\mathrm{Mpc}$ to
    $r_p{=}30.0\,h^{-1}\,\mathrm{Mpc}$ and 20 linearly-spaced bins of
    $p(N_8^g)$ from $N_8^g{=}0.0$ to $N_8^g{=}160.0$.}
\label{fig:cova}
\end{figure}

We use the jackknife resampling technique to compute covariance matrices
for our observables. In our analysis we utilize measurements from the SDSS
galaxy sample as well as a mock sample derived from the ELUCID simulation.
In either case we compute two jackknife re-sampled covariance matrices, one
for $w_{p}$ and $p(N_8^g)$ and the other for the five $N_8^g$-selected
correlation functions $w_{p} (r_p | \tilde{N}_8^{g,i})
\rvert_{i{=}1\mbox{-}5}$. Using multiple realizations of our ELUCID mock
sample we have computed the contribution to the covariance from using a
single random realization of the HOD. We have found that this realization
contribution to the error is strongly subdominant at all scales for all of
our observables and is therefore ignored.

The covariance matrices computed on different mocks and the SDSS catalog
generally exhibit similar behaviors. However, it should be noted that
quantitatively there are important differences. The sample variance for our
observables will obviously scale with their signal, but the presence or
absence of assembly bias will also have an effect. Two mocks with the same
signal but different values of our assembly bias parameters will have
different observable covariance matrices. If we consider the example of
$w_p$, we see that these two mocks will have different covariances because
assembly bias is adding to the spatial variation of $w_p$ due to the large
scale density. In the case of the distribution of $N_8^g$, the effect is
even more direct because any galaxy assembly bias acts to boost the
variance of $N_8^g$.

\begin{figure*}
\centering \includegraphics[width=1.0\textwidth]{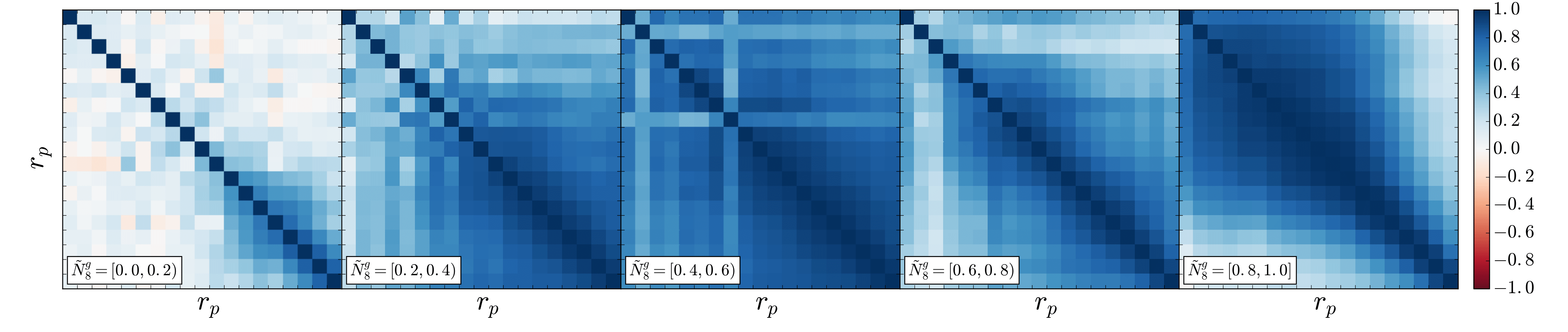}
\caption{Correlation matrix of the projected cross-correlation function
between galaxy quintiles selected by $N_8^g$ and the overall galaxy sample,
$w_p (r_p | \tilde{N}^{g,i}_8)\rvert_{i{=}1\mbox{-}5}$, with the
correlation
    coefficients colour-coded by the colour bar on the far right. In each
    panel bins from left (top) to right (bottom) correspond to 20 logarithmically-spaced
    bins of $w_p(r_p|\tilde{N}^{g,i}_8)$ from
    $r_p{=}0.3\,h^{-1}\,\mathrm{Mpc}$ to
    $r_p{=}30.0\,h^{-1}\,\mathrm{Mpc}$.} \label{fig:covb}
\end{figure*}

The correlation matrices measured in SDSS for $\{w_{p}, p(N_8^g)\}$ and
$w_{p} (r_p | \tilde{N}_8^{g,i})\rvert_{i{=}1\mbox{-}5}$ are shown in
Figure~\ref{fig:cova} and \ref{fig:covb}, respectively. Turning towards
Figure \ref{fig:cova} we see that for $w_{p}$ there are significant
correlations among bins within the 1-halo regime and bins within the 2-halo
regime, but there is little correlation between bins across these two
regimes. For $p(N_8^g)$ we see that the covariance is mostly diagonal with
minor correlations between nearby bins. We also see that there are negative
correlations between large and small $N_8^g$ bins. This reflects the
explicit integral constraint on the probability distribution function.

Overall, the covariance between the two types of observables is weak but
exhibits some structure. There are negative covariances between all bins of
$w_{p}$ and bins of low $N_8^g$ and positive covariance between all bins of
$w_{p}$ and bins of high $N_8^g$. This is because galaxies with high values
of $N_8^g$ are in high density regions and therefore have stronger clustering,
while those with low values of $N_8^g$ are in low density regions and have
weaker clustering.

Figure \ref{fig:covb} shows the correlation matrix of our five $N_8^g$
selected correlation functions. Going from left to right, the correlation
matrix for increasing quintiles of $N_8^g$ starts with the bottom quintile
on the far left and ends with the top quintile on the far right. In all
cases we see roughly the same structure we observe for the global $w_{p}$,
there are significant correlations between nearby bins but little
correlation between large and small scales. Additionally, the low-quintile
correlation functions are more diagonally dominated than the high-quintile
correlation functions. This is because the shot-noise component, which is
diagonal and constant across the five quintiles, becomes more dominant when
the sample-variance component, which scales with the signal, is weaker.
Using jackknife resampling we have also calculated the covariance between
our five $N_8^g$ selected cross-correlation functions. We have found these
errors to be extremely noisy and near zero. Therefore we chose to set them
equal to zero in what follows.

\section{Model Tests with Standard Halo Occupation Distribution}
\label{app:noGABtest}

\begin{table*}[hbt!]
\renewcommand*{\arraystretch}{1.3}
   \centering \caption{Posterior constraints of model parameters from an
   ELUCID mock catalog with significant levels of galaxy assembly bias used
   as data. Likelihood's are calculated using ELUCID and Bolshoi haloes with $\Qcen$ and $\Qsat$ fixed to zero.  Values quoted are posterior modes with $68\%$ confidence intervals.  The value of $\chi^2 / \mathrm{d.o.f}$ is calculated from the mean of the posterior samples. Input values are listed in the last row.}
    \begin{adjustbox}{center}
    \begin{tabular}{llccccccccc}
    \hline
    Data vector & Haloes & $\siglogM$ & $\log M_\mathrm{min}$ & $\log M_1$ & $\alpha$ & $\Qcen$ & $\Qsat$ & $\Acon$ & $\chi^2 / \mathrm{d.o.f}$ \\
    \hline
    $\{w_p(r_p), n_g \}$ & ELUCID & $0.20^{+0.17}_{-0.13}$ & $12.00^{+0.06}_{-0.05}$ & $13.00^{+0.11}_{-0.12}$ & $0.90^{+0.09}_{-0.08}$ & - & - & $0.85^{+0.12}_{-0.12}$ & $13.01/16$\\
    $\{w_p(r_p), n_g \}$ & Bolshoi & $0.19^{+0.17}_{-0.13}$ & $12.01^{+0.06}_{-0.06}$ & $12.80^{+0.11}_{-0.12}$ & $0.80^{+0.08}_{-0.08}$ & - & - & $0.87^{+0.12}_{-0.12}$ & $42.23/16$\\
    \hline
    $\{w_p(r_p), p(N_8^g), n_g \}$ & ELUCID & $0.26^{+0.10}_{-0.10}$ & $12.03^{+0.06}_{-0.05}$ & $12.79^{+0.08}_{-0.09}$ & $0.80^{+0.06}_{-0.06}$ & - & - & $0.81^{+0.13}_{-0.12}$ & $36.99/36$\\
    $\{w_p(r_p), p(N_8^g), n_g \}$ & Bolshoi & $0.19^{+0.12}_{-0.11}$ & $11.97^{+0.05}_{-0.04}$ & $12.72^{+0.07}_{-0.07}$ & $0.79^{+0.05}_{-0.05}$ & - & - & $0.86^{+0.12}_{-0.12}$ & $54.51/36$\\
    \hline
    $\{ w_{p} (r_p | \tilde{N}^{g,i}_8)\rvert_{i{=}1\mbox{-}5}, n_g \}$ & ELUCID & $0.16^{+0.06}_{-0.07}$ & $12.03^{+0.03}_{-0.03}$ & $12.96^{+0.08}_{-0.09}$ & $0.85^{+0.05}_{-0.05}$ & - & - & $0.74^{+0.12}_{-0.14}$ & $65.38/96$\\
    $\{ w_{p} (r_p | \tilde{N}^{g,i}_8)\rvert_{i{=}1\mbox{-}5}, n_g \}$ & Bolshoi & $0.17^{+0.07}_{-0.07}$ & $12.01^{+0.03}_{-0.03}$ & $12.81^{+0.07}_{-0.09}$ & $0.79^{+0.04}_{-0.05}$ & - & - & $0.69^{+0.14}_{-0.15}$ & $141.51/96$\\
    \hline
    $\{ w_{p} (r_p | \tilde{N}^{g,i}_8)\rvert_{i{=}1\mbox{-}5}, p(N_8^g), n_g \}$ & ELUCID & $0.20^{+0.07}_{-0.07}$ & $12.00^{+0.03}_{-0.02}$ & $12.84^{+0.05}_{-0.06}$ & $0.83^{+0.04}_{-0.04}$ & - & - & $0.71^{+0.14}_{-0.14}$ & $96.57/116$\\
    $\{ w_{p} (r_p | \tilde{N}^{g,i}_8)\rvert_{i{=}1\mbox{-}5}, p(N_8^g), n_g \}$ & Bolshoi & $0.18^{+0.08}_{-0.08}$ & $11.96^{+0.03}_{-0.02}$ & $12.75^{+0.04}_{-0.05}$ & $0.81^{+0.03}_{-0.04}$ & - & - & $0.69^{+0.15}_{-0.16}$ & $175.98/116$\\
    \hline
    Input Value & - & 0.20 & 11.95 & 13.15 & 1.0 & -0.50 & 0.50 & 0.86 & - \\
    \hline
  \end{tabular}
  \end{adjustbox}
\label{table:app_noGAB_mock_constraints}
\end{table*}

In section \ref{sec:mock} we performed a variety of mock tests using two
ELUCID mocks, one without and the other with strong galaxy assembly bias.
During those mock tests, we allow $\Qcen$ and $\Qsat$ to vary freely.
For the sake of completeness here we
include analogous mock tests
by fitting a standard HOD model with fixed values of $\Qcen=\Qsat=0$
to the ELUCID mock with strong galaxy
assembly bias~($\Qcen=-0.50$, $\Qsat=0.50$). The results of these mock
constraints are listed in Table \ref{table:app_noGAB_mock_constraints}; values
quoted correspond to parameter posterior modes and $68\%$ confidence intervals. Input values for the HOD parameters are listed in the final
row of the table.

\end{appendix}

\label{lastpage}

\end{multicols}
\end{document}